\documentclass{elsarticle}

\usepackage{hyperref}

\usepackage[utf8]{inputenc}
\usepackage{graphicx}
\usepackage{float}
\usepackage{placeins}
\usepackage{xcolor}
\usepackage{amsmath}
\usepackage{amssymb}

\usepackage{caption}

\newcommand{\JS}{\textcolor{black}}

\begin{document}

\begin{frontmatter}

\title{Design and characterization of an instrumented slider aimed at measuring local micro-impact forces between dry rough solids}

\author[1]{C. Grégoire}
\author[2]{B. Laulagnet}
\author[1]{J. Perret-Liaudet}
\author[1]{T. Durand}
\author[1]{M. Collet}
\author[1]{J. Scheibert}
\ead{julien.scheibert@ec-lyon.fr}

\address[1]{Univ Lyon, Ecole Centrale de Lyon, ENISE, ENTPE, CNRS, Laboratoire de Tribologie et Dynamique des Syst\`emes LTDS, UMR 5513, F-69134, Ecully, France}
\address[2]{Laboratoire Vibrations Acoustique, INSA Lyon, 25 bis avenue Jean Capelle 69621 Villeurbanne Cedex, France}
\address{\textnormal{Published in \href{https://doi.org/10.1016/j.sna.2020.112478}
	{Sensors and Actuators: A. Physical 317:112478, 2021}}}

%
%
%
%

\begin{abstract}
Sliding motion between two rough solids  under light normal loading involves myriad micro-impacts between antagonist micro-asperities. Those micro-impacts are at the origin of many emerging macroscopic phenomena, including the friction force, the slider's vibrations and the noise radiated in the surroundings. However, the individual properties of the micro-impacts (\textit{e.g.} maximum force, position along the interface, duration) are essentially elusive to measurement. Here, we introduce an instrumented slider aimed at measuring the position and the normal component of the micro-impact forces during sliding against a rough track. It is based on an array of piezoelectric sensors, each placed under a single model asperity. Its dynamical characteristics are established experimentally and compared to a finite elements model. We then validate its measurement capabilities by using it against a track bearing simple, well-defined topographical features. The measurements are interpreted thanks to a simple multi-asperity contact model. Our slider is expected to be useful in future studies to provide local insights into a variety of tribological questions involving dry rough sliding interfaces.
\end{abstract}

\begin{keyword}
Array of piezoelectric sensors \sep Local force measurement \sep  Tribology \sep Sliding rough contact  \sep Roughness noise
\end{keyword}

\end{frontmatter}


\section{Introduction}
\label{intro}
 
Contacts between solids are submitted to complex force fields developing along the interface, the characteristics of which depend, among others, on the external loading, the macroscopic geometry and the surface topographies (see e. g.~\cite{vakis_modeling_2018} for a recent review). Because those force fields directly control several key tribological phenomena including energy dissipation, wear or sliding-induced vibrations, the measurement of local contact forces is highly desirable. However, in general, such measurements remain challenging due to the difficulty to place non-invasive local force sensors in the close vicinity of the interface. Still, with the goal of better understanding the elementary mechanisms occurring along frictional interfaces, a number of local force measurement methods have been proposed in the literature.

One strategy is to embed one or several force sensors inside the bulk of one of the two solids in contact. Each sensor, with a lateral size $l$, probes the stress field at a certain depth $h$ below the surface. Those non-vanishing length scales are responsible for the limited spatial resolution of the measurement: first, the local stress at a depth $h$ is an average over the surface stresses within a region of typical extension $h$ ; second, the stresses at depth are themselves integrated over the sensor area. Overall, one expects the sensor output to be sensitive to the interfacial forces acting in a region of typical lateral extension $l+h$. Various technologies have been tested for the embedded sensors, including strain gauges~\cite{svetlizky_brittle_2019}, piezoelectric~\cite{howe_dynamic_1993} or piezoresisting elements~\cite{jung_piezoresistive_2015,okatani_tactile_2016}, pressure-sensitive electric conductive rubber (PSECR)~\cite{zhang_embedded_2013} and MEMS~\cite{candelier_role_2011}, with applications not only to tribology, but also to haptic sensors (see~\cite{dahiya_robotic_2013, saccomandi_microfabricated_2014} for reviews). For instance, millimeter-sized MEMS forces sensors embedded at the base of a rough elastomer slab of millimetric thickness have been used not only to investigate the pressure and shear stress fields at the contact between the rough slab and rigid smooth sliders~\cite{scheibert_experimental_2008,scheibert_stress_2009}, but also to unravel the effect of fingerprints on the tactile perception of fine textures~\cite{scheibert_role_2009}.

Another strategy is to probe the rough contact interface in a minimally-invasive way, for instance using an imaging technique through a transparent material. The shear stress field can be accessed by inversion of the in-plane displacement field~\cite{chateauminois_local_2008,prevost_probing_2013}, which can be obtained by following the motion of appropriate tracers, either incorporated on purpose~\cite{chateauminois_local_2008,lengiewicz_finite_2020} or naturally present in the image~\cite{prevost_probing_2013}. The spatial resolution of such measurements is typically limited by both the depth of the tracers and their inter-distance (although the latter limitation can be significantly reduced in steady state conditions~\cite{chateauminois_local_2008}), or by the size of the correlation box when a digital image correlation (DIC)-like method is used to measure the displacements~\cite{prevost_probing_2013}. If the surface topography is specifically engineered as a population of well-defined micro-spheres, both shear and normal forces on each micro-contact can be estimated from its in-plane displacement and true contact area~\cite{romero_probing_2014}.

In this work, our objective is to propose a measurement tool useful to better understand the phenomena occurring at the sliding interface between two rough metallic surfaces under light normal loading. In such lightly loaded interfaces, the contact pressure is much smaller than the materials' elastic moduli, so that each micro-contact induces local displacements which remain negligible compared to the size of the largest asperities in the topographies. In those conditions, the interactions between two rough surfaces is expected to be essentially geometrical and to be localised at the (almost undeformed) tip of a few individual asperities at each instant. The sliding motion of a rough slider on a rough track thus induces a motion normal to the interface, as the slider passes over a succession of random asperities of the track~\cite{ponthus_statistics_2019}.

When the slider and/or track are not perfectly rigid, and when the sliding speed is large enough, the interaction between the two rough surfaces is a succession of micro-impacts between antagonist asperities, which trigger a vibration of the solids, through not only their rigid body modes, but also their other eigenmodes. Overall, those vibrations are the source of the so-called roughness noise~\cite{akay_acoustics_2002}, the empirical laws of which have been studied extensively~\cite{yokoi_fundamental_1982,stoimenov_roughness_2007,le_bot_measurement_2010,ben_abdelounis_experimental_2010}. However, such empirical laws remain unexplained, and both numerical modelling~\cite{dang_direct_2013} and statistical analyses~\cite{le_bot_noise_2017} point towards the challenging need to better describe the forces involved in the individual, sub-millisecond-lived micro-impacts.

\begin{figure}[htb!]
\centering
\includegraphics[width=0.75\linewidth]{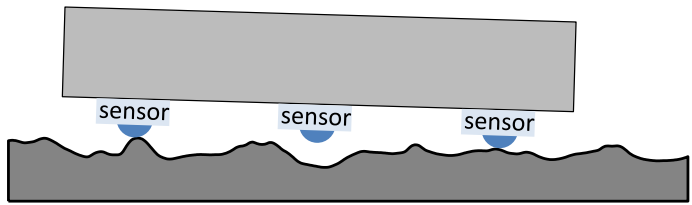}
\caption{Sketch of an instrumented slider (top) on a rough track (bottom). The slider's surface is equipped with several model asperities (blue spherical caps), each of them being monitored by a dedicated local force sensor.}
\label{SchView}
\end{figure}

To address this challenge, none of the two above-mentioned measurement strategies (embedded sensors and optical monitoring) is suitable. First, metals being opaque to light, imaging methods cannot be applied. Second, the non-vanishing size of the region probed by a sensor buried inside the solid material generally encompasses a large number of surface asperities, thus impeding identification of individual asperity/asperity impacts. Here, to overcome these difficulties, our strategy is to design a slider equipped with an array of well-defined surface asperities, each of them being monitored by a dedicated force sensor (Fig.~\ref{SchView}). The expected short duration of micro-impacts, typically shorter than a millisecond~\cite{dang_direct_2013}, requires a large measurement bandwidth to be time-resolved, leading us to choose piezoelectric elements as sensors.

In this paper, our scope is not to address directly the problem of the roughness noise and sliding-induced vibrations under light normal loading, but to present a new \textit{in situ} measurement tool which will be useful for that purpose in future works. The paper is organised as follows. First, we describe the main elements of conception of the slider, its practical realisation, and its calibration (section~\ref{sec:slider}). Then, we demonstrate its tribological relevance by using it on a track with simple topographical features (section~\ref{sec:ExpResults}), and by interpreting the signals with a simple multi-asperity model (Supplementary Material, SM).

\section{Description and calibration of the slider}
\label{sec:slider}
Before describing the details of the developed instrumented slider (section~\ref{sec:sliderSpec}), let us note that general micro-impact forces are expected to have random directions with respect to the interface. Indeed, the tangent plane of each micro-contact can very well be different from the average plane of the macroscopic contact interface. In addition, each micro-contact is submitted not only to repulsion forces normal to the local contact tangent plane to avoid interpenetration of the solids, but also to friction forces acting along the local contact tangent plane. Thus, one would ideally want to access the three components of the micro-contact forces, which would require three co-localised sensors associated to each model asperity at the slider's interface. In this first work, to avoid an excessive complexity in the design and realisation of the slider, we decided to use only one sensor unit per asperity, to measure only the projection of the micro-impact forces on the normal to the slider's average plane.

One could imagine that the slider may simply be prepared with an array of model asperities, each of which being directly attached to a dedicated piezoelectric element (like in Fig.~\ref{SchView}). However, during sliding on a rough frictional interface, the piezoelectric elements could be submitted to non-negligible tangential forces, with two different undesired consequences. First, a given piezoelectric element is generally sensitive to all directions of an external stimulus (although with different sensitivities), making it impossible to separate the different contributions from a single charge output~\cite{gautschi_piezoelectric_2002}. Second, piezoelectric ceramics can easily be damaged by an even modest shear stimulus~\cite{watchman_mechanical_2009}. Thus, a large part of the chosen design will be justified by the necessity to expose the piezoelectric elements to normal forces only. In practice, we have tested various design options, each of which has been evaluated using a finite element model (FEM) of the full instrumented slider (section~\ref{sec:fem}). The final design has thus been selected after a FEM-assisted trial-and-error procedure. The predicted capabilities of the slider, in particular in terms of locality and bandwidth of the measurements, are furthermore affected by the rest of the measurement chain (section~\ref{sec:instrum}), and are finally compared to actual measurements and calibrations in section~\ref{sec:calibration}. The signal analysis allowing efficient localisation and force estimates of individual micro-impacts is discussed in section~\ref{sec:localization}.

\subsection{Slider's specifications and mechanical assembly}\label{sec:sliderSpec}
\begin{figure}[htb!]
\centering
\includegraphics[width=0.8\linewidth]{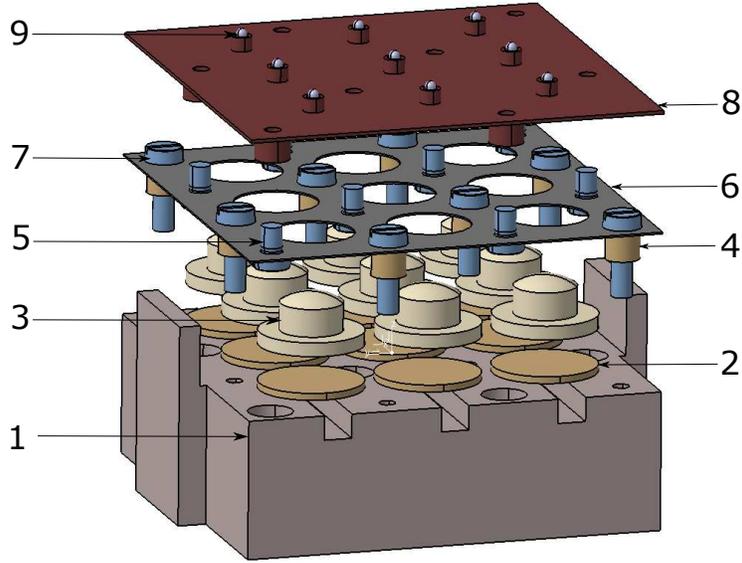}
\caption{Exploded 3D view of the various mechanical parts constituting the instrumented slider. 1: Aluminum base. 2: Piezoelectric elements. 3: Sphere-ended aluminum fitting parts. 4: Hollow cylinder-shaped spacer. 5 \& 7: M3 screws. 6: Hollow carbon steel spring plate. 8: Carbon steel top plate, featuring 8 threaded hollow cylinders on its bottom, and 9 sphere holders on its top. 9: \JS{Bearing steel sphere}.}
\label{fig:exploded}
\end{figure}
Figure~\ref{fig:exploded} (bottom right) shows an exploded view of all mechanical parts constituting the instrumented slider. The sensing units (label 2 in Fig.~\ref{fig:exploded}) are nine circular plate-shaped elements (15\,mm diameter, 1\,mm thick) of a piezoelectric ceramic (PZ27, Ferroperm F1270508) with the electrodes on the flat surfaces. The piezoelectric elements are first glued with an epoxy glue at the surface of a thick anodised aluminum (2017A) base (label 1). On top of each of them is then glued a fitting part (label 3) having a cylindrical base with the same diameter as the piezoelectric element, topped by a cylindrical body with a smaller diameter, itself ended by a spherical cap (radius of curvature 7.5\,mm). Precise in-plane positioning of the piezoelectric elements and their associated fitting parts during gluing was achieved using a dedicated 3D-printed positioning part. To minimise the thickness of the glue films, gluing was performed under a dead weight of 2.6\,kg. The resulting altitudes of the summits of the nine spherical caps was measured with an interferometric profilometer, and found to deviate from their mean plane by less than 46\,$\mu$m each (32\,$\mu$m standard deviation).

The top surface of the slider (on which the model asperities will be attached), is a carbon steel plate of thickness 0.5\,mm (label 8). It is pressed onto the nine spherical caps (label 3) using a thinner (0.25\,mm thick) carbon steel spring plate (label 6) featuring nine holes through which the spherical caps can pass and come into contact with the top plate. The spring plate is first mounted on the base (label 1) through eight top-headed screws (label 7), each passing through a cylindrical spacer (label 4). The top plate is then attached to the spring plate with eight down-headed screws (label 5) screwed into eight threaded cylinders spot-welded on the bottom surface of the top plate. Those screws can be actuated from the bottom surfaces of the base, through eight dedicated through-holes.

The spring plate serves two crucial roles. First, its relatively small bending stiffness is used to pull the top plate in contact with all nine spherical caps, although they have slightly different altitudes. In practice, the design of Fig.~\ref{fig:exploded} includes an initial 300\,$\mu$m gap between the spring plate and the threaded cylinders attached to the top plate. So, by fine tuning the angles of all eight individual down-headed screws (label 5), a pre-load can \JS{in principle} be applied and adjusted on each spherical cap. \JS{Here, we did not use this fine-tuning capability and fully screwed all down-headed screws. Thus, no progressive loosening is expected during the lifetime of the slider, so that no additional anti-unscrewing glue was deemed necessary.} The pre-load is chosen large enough so that the contact between spherical caps and top plate is never lost, even in the case of transient negative loads, for instance during potential vibrations of the top and/or spring plates. The sphere/plane geometry of the contact imposes a well-defined location of the normal force on the fitting part, which, for a pure normal load applied by the top plate, eliminates any torque on the surface of the piezoelectric element. Second, the large in-plane stiffness of the spring plate essentially prevents any lateral motion of the top plate relative to the spherical caps, even in case a significant shear load is applied to the top plate. The spring plate is thus the main design element ensuring that the piezoelectric sensors are exposed to almost purely normal stimuli.

The top surface of the top plate further features nine spot-welded cylinders with an upper conical hole, into which model asperities can be fixed. The base also features two lateral arms, which will prove useful to push the slider during tribological experiments (section~\ref{sec:ExpResults}). The grooves on the top surface of the base serve as channels to fit the cables connecting the nine piezoelectric elements, as described in section~\ref{sec:instrum}. The fact that the piezoelectric elements are glued above the grooves and not between two perfectly flat surfaces may cause a non-linearity of the sensors, but we have checked that, in our case, this effect remains negligible (see section~\ref{sec:fem}). All dimensions of the slider can be found in the technical drawings of Fig.~S1 (in SM).

\begin{figure}[htb!]
\centering
\includegraphics[width=0.75\linewidth]{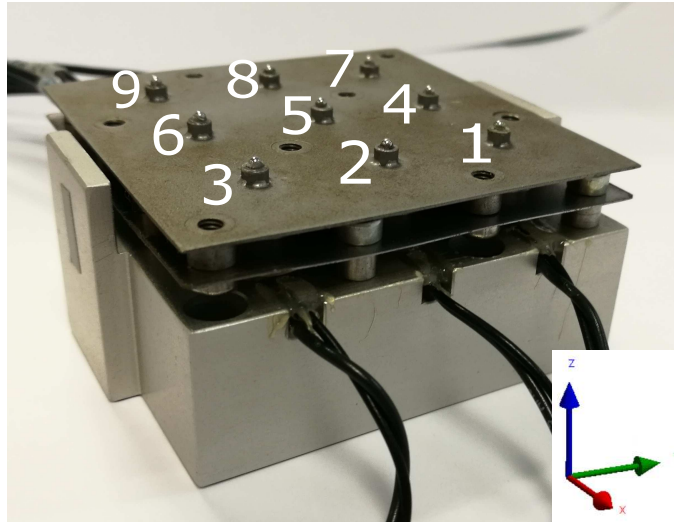}
\caption{View of the slider indicating the labels of all asperities, used all along the article.}
\label{Pad}
\end{figure}

Finally, \JS{a bearing steel sphere (type SAE 52100, particularly resistant to wear)} with a radius of curvature of \JS{0.75}\,mm is glued in each conical hole of the top plate, so that the slider features nine spherical asperities at its surface. A picture of the complete slider is shown in Fig.~\ref{Pad}. In the conditions used in the rest of this study, all contact interactions between the slider and a track will only occur through one or several of those nine potential spherical asperities. Their individual altitudes have been measured with an interferometric profilometer, and are provided in Table~S1 (in SM) with respect to their mean plane. Their standard deviation is 72$\mu$m while their maximum difference is
203$\mu$m.

\subsection{Finite element analysis}
\label{sec:fem}
To design the instrumented slider, we built several finite element (FE) models of the slider, with different architectures and dimensions. Here, we present the static and modal characteristics predicted by the FE model of our final design solution, described in section~\ref{sec:sliderSpec}.

\begin{figure}[htb!]
\centering
\includegraphics[width=0.8\textwidth]{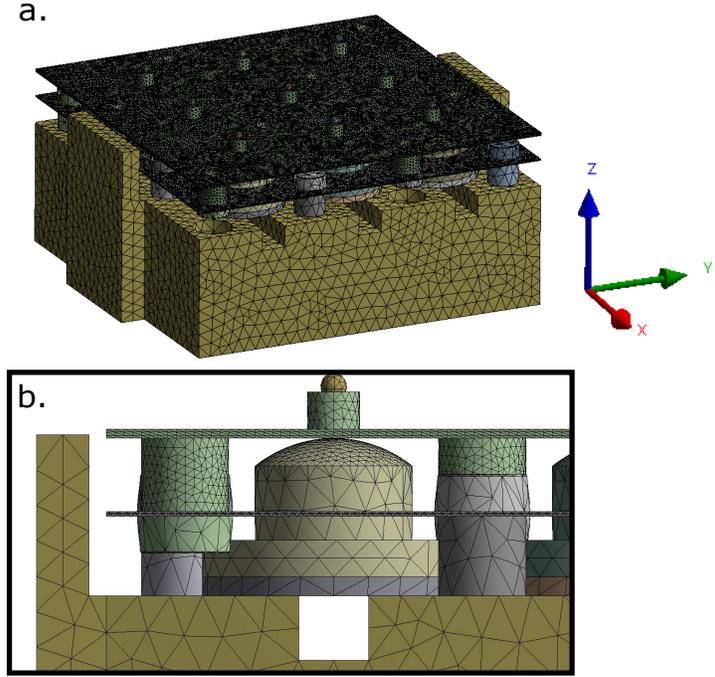}
\caption{Mesh of the slider. a: full mesh. b: zoom on one sensing unit (label 3 in Fig.~\ref{Pad}).}
\label{fig:mesh}
\end{figure}

The entire instrumented slider was discretized and analysed using the software Ansys. As shown in Fig.~\ref{fig:mesh}, all parts of the slider were discretized into tetrahedral 3-D elements. They possess 10 nodes, 3 degrees-of-freedom per node corresponding to translations in the 3 directions, and such that displacements have quadratic approximations. Elements are used with homogeneous and isotropic elastic solid behaviour, except for the piezoelectric sensors, which are anisotropic. The material properties of all materials are reported in Tab.~S2. The total final number of elements is up to 340000, leading up to 610000 nodes and about 1.8\,million degrees-of-freedom. The most critical part of the model is the discretization of the spring plate (label 6 in Fig.~\ref{fig:exploded}), because a minimum number of elements in the thickness is required. To determine this minimum number, we have performed two preliminary convergence tests. The first test is based on the modal analysis of a simply supported, 0.25\,mm thick, \JS{60\,mm $\times$ 60\,mm} steel plate, for which natural frequencies and modal shapes are analytically known. A good agreement was obtained with a 0.1\,mm element size, leading to relative errors less than 0.2\% for the first ten natural frequencies, and to well-reproduced modal shapes. The second test consisted in the convergence of natural frequencies of a 0.25\,mm thick, \JS{60\,mm $\times$ 60\,mm} free-free plate with the same holes as in our final design. Again, a 0.1\,mm element size was found adequate. Thus, we used 0.1\,mm-sized elements to discretize both the top plate (label 8, 147000 elements) and the spring plate (label 6, 66000 elements), while 2\,mm elements were used for the base (label 1, 70000 elements) and for each sensor (label 2, 350 elements), long screws (label 7, 1800 elements), small screws (label 5, 1300 elements) and spacer (label 4, 200 elements). For the spherical-capped fitting parts (label 3, 3300 elements), the element size was 2\,mm except for the cap, which had an element size of 0.5\,mm.
Junctions between parts are modelled as fully bonded contacts, except for the connection between the top plate and spherical caps. For the latter, we used unilateral normal contact and Coulomb friction with a friction coefficient of 0.5, a conservative value larger than the expected range for aluminum/steel contacts~\cite{muller_formulaire_2000}.

We evaluated the relationship between a static load applied to each surface asperity and the resulting force in all nine sensors. For such static finite element calculation, we assumed that the base is perfectly rigid. So, we removed it from the model and instead imposed clamped boundary conditions on the corresponding faces of the other parts usually in contact with the base. The contact between the top plate and each of the nine spherical caps was pre-loaded in a similar way as in the assembly of the experimental slider (see section~\ref{sec:sliderSpec}): the 300\,$\mu$m air gap between the spring plate and  the threaded cylinders of the top plate was first replaced by the same volume of steel ; then all top-headed screws and spacers (labels 7 and 4 in Fig.~\ref{fig:exploded}) were displaced downward by \JS{a common value of} 300\,$\mu$m. The resulting deformations  in the spring plate have an amplitude of 300\,$\mu$m, while the (stiffer) top plate deforms by no more than about 30\,$\mu$m. The generated pre-loads were about 16\,N on sensors 1, 3, 7 and 9 \JS{(the sensors on the four corners of the top plate)}, 12\,N on sensors 2, 4, 6, 8 \JS{(on the four sides)} and 7\,N on sensor 5 \JS{(on the center)}. For each surface asperity $i$, a simulation with an additional vertical force $F_i$=10\,N applied on it was carried out, and the transmitted force in all sensors was calculated. The results were summarised as a static response matrix $\mathbf{H}$ (Fig.~S2). Each element $H_{ij}$ represents the resulting vertical force on the ${j^{th}}$ sensor induced by the force applied on the ${i^{th}}$ spherical cap. $\mathbf{H}$ is a strictly diagonally dominant matrix, with $ \sum_{j\neq i}{\left| H_{ij}\right|} < 0.05 {\left|H_{ii}\right|}$ and $ H_{ii} > 0.95 F_{i} $.
In conclusion, a static force applied on a given surface asperity is transmitted to the corresponding sensor with less than 5\% rejection on the other sensors, which was considered close enough to the desired behaviour for our final design.

Similarly, we applied tangential loads between 1 and 100\,N on each of the surface asperities, and looked at the induced tangential forces in all piezoelectric sensors. The latter forces have been found to be always smaller than 9\,\% of the applied load. The equivalent shear stress was always below 0.1\,MPa, well below the yield strength of the piezoelectric sensors (about 40\,MPa). This result indicates that our design is expected to be effective in protecting the piezoelectric elements from potentially damaging shear forces.

To test the potential influence of the grooves above which the piezoelectric sensors are glued (see Fig.~\ref{fig:mesh}(b)), we performed the following additional static FE simulations, using the extension ACT Piezo \& MEMS of Ansys. We considered a single sensing unit made of one piezoelectric element (label 2 in Fig.~\ref{fig:exploded}, piezoelectric moduli $e_{31}$=-3.09\,C.m$^2$, $e_{33}$=16.0\,C.m$^2$, $e_{15}$=11.64\,C.m$^2$, and relative permittivities at constant strain $\epsilon^S_{1,r}$=1129.69 and $\epsilon^S_{3,r}$=913.73, equipotential condition on both faces) and its spherically-capped fitting part (label 3), bonded either on a flat solid (reference case) or on a grooved solid. In both cases, a nominal normal force was applied over a circle of radius 1\,mm around the apex of the spherical cap, together with an additional tangential load (along $x$ or $y$) in the range [0--0.1]\,N and/or an additional torque (around $x$ or $y$) in the range [0--0.1]\,N/m. In all those cases, the difference between the reference and grooved cases was found less than 3.6\% for the electric potential difference (voltage) generated between the two faces of the piezoelectric element, suggesting a negligible influence of the groove on the force measurement.

To qualify the frequency domain on which the instrumented slider conveniently responds, and its dynamic performance, we then performed a (purely mechanical, i.e. not including piezoelectricity) modal analysis for the full model (i.e. now including the base) over the frequency range [0-10]\,kHz. The bottom face of the base has fully constrained displacements, and, to ensure linearity of the calculations, the contacts between sphere-ended fitting parts and top plate are now fully bonded over a circle of radius 0.47\,mm. The eigenshapes of the two first eigenmodes (eigenfrequencies in the range 1.5--1.9\,kHz) are shown in Fig.~S3. Each eigenshape consists of a particular combination of deformations of the top and spring plates. Modal displacements at the locations of the nine contacts remain very small, due to the high vertical stiffness of the piezoelectric ceramic and fitting part compared to that of the top and spring plates.

\begin{figure}[htb!]
\centering
\includegraphics[width=0.9\textwidth]{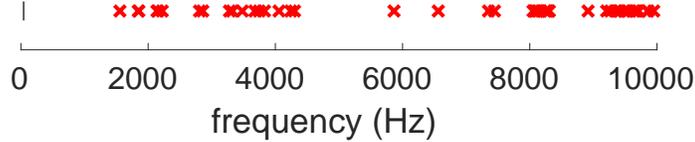}
\caption{Eigenfrequencies of the slider up to 10\,kHz, as calculated with the FE model with the bottom face fixed.}
\label{fem}
\end{figure}

Figure~\ref{fem} presents all the calculated eigenfrequencies. As can be seen, they are not evenly distributed over the frequency range, but form several mode packs. For plates in flexural vibration, the modal density is expected to be constant versus frequency~\cite{lebot_foundation_2015}. However, mode packing arises when the system has a sort of spatial periodicity~\cite{laulagnet_sound_1990}, as is the case with the regular arrangement of connections between the top and spring plates, and between the spring plate and base. Such a design is thus presumably responsible for the observed heterogeneous modal density.

The first mode pack is found at around [1.5--2.2]\,kHz, so that we can assume a static behaviour of the instrumented slider in the range [0--1.5]\,kHz. From this point of view, the adopted design is expected to enable monitoring of impacts of duration about and below the millisecond, one of the desired features mentioned in the introduction. Based on the static results, we then expect that, below 1.5\,kHz, the rejection on sensor $ {j\neq i} $ of an impact occurring on asperity $i$ will remain less than about 5\%.

\begin{figure}[htb!]
\centering
\includegraphics[width=0.99\columnwidth]{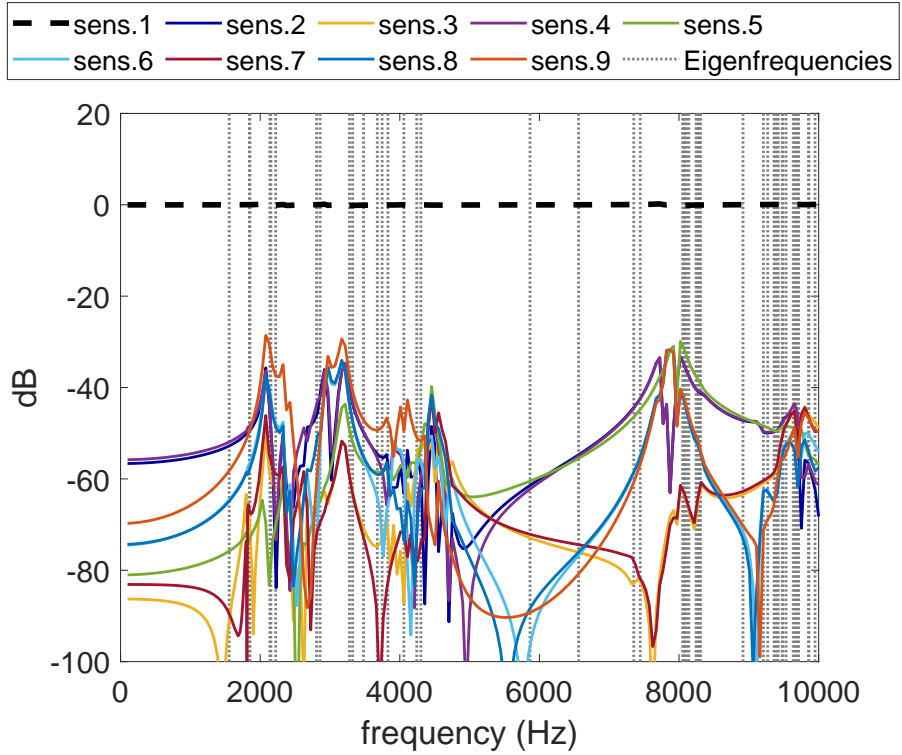}
\caption{Force frequency response of all sensors when asperity 1 is excited, in dB ($20log_{10}\left[\frac{\text{Force response}}{\text{Excitation}}\right]$). Dotted vertical lines: eigenfrequencies predicted by the FE modal analysis.}
\label{fig:Harmonic}
\end{figure}

To better assess the expected dynamical behaviour of the slider, we finally calculated the force frequency response of all sensors to a harmonic vertical stimulus on a single asperity, using the mode superposition method. The results are shown on Fig.~\ref{fig:Harmonic} in the case of a stimulus on asperity 1. As expected, the force frequency response of sensor 1 is very close to 1 for all frequencies, because the force on asperity 1 is directly transmitted to sensor 1 with little loss through the other sensors. Indeed, below 1.5\,kHz, the force frequency responses of all other sensors are smaller than -55\,dB. Interestingly, the two sensors with the largest response are sensors 2 and 4, which are the nearest neighbours of asperity 1. The responses are actually frequency-dependent, and in particular peaks can be seen in the vicinity of the eigenfrequencies, especially around 2, 3 and 8\,kHz. Nevertheless these peaks remain small (maximum value of -30\,dB), meaning that, even in the highly dynamic frequency domain, the excited sensor is still predicted to capture most of the imposed dynamical force.

\subsection{Related instrumentation}\label{sec:instrum}
In order to acquire the time-evolution of the nine pressure signals, a shielded cable (AC-0005-K, Brüel \& Kj{\ae}r) was first welded on each piezoelectric element, before the gluing step described in section~\ref{sec:sliderSpec}. Both welding points are located on the same face of the piezoelectric plate-shaped elements, thanks to one of the electrodes being wrapped around the thickness of the element. Thus, for the gluing step, the extra thickness due to the two welding points and the emerging cables can be fitted into the grooves present on the top surface of the base, \JS{toward the outside of the slider (see the cables emerging from the channels in Fig.~\ref{Pad})}.

The cable conveying the charges generated by each of the nine piezoelectric elements is connected to one input channel of a conditioning amplifier (2694A, Brüel \& Kj{\ae}r, bandwidth [1\,Hz--50\,kHz]), through a charge-to-DeltaTron converter (2647, Brüel \& Kj{\ae}r, 1\,mV/pC, bandwidth [0.17\,Hz--50\,kHz]). The outputs of the conditioning amplifier are finally acquired using a 32-channels recorder and analyser (OR38, OROS). The acquisition rate is set to 25\,kHz, ensuring that frequencies up to 12.5\,kHz are adequately acquired. Note that the overall bandwidth of the measurement chain is [1--12500]\,Hz, which implies that the average value of the pressure signal is not available. In the dynamical conditions of a rough/rough contact sliding at sufficiently high speeds and involving impacts in the millisecond range, as targeted with this instrumented slider, this is not a limitation. However, for combined low sliding speeds and large wavelengths of the topography, like in some of the test conditions explored in section~\ref{sec:resultsDefects}, the lower-frequency, information-bearing contents of the pressure signals may be filtered out from the outputs.

The multi-channels acquisition device is used not only for the pressure outputs from the piezoelectric sensors, but also for the other measured signals: impact-hammer force (section~\ref{sec:calibration}) and friction force (section~\ref{sec:ExpResults}), so that all of them are acquired with the very same time frame.

\subsection{Dynamical characterisation}\label{sec:calibration}
To calibrate and characterise the dynamical capabilities of our instrumented slider, we conducted a series of experiments in which the slider was first clamped to the table. Then each of its nine individual surface asperities was submitted to short stimuli by an impact hammer (PCB piezotronics, model 086C03) equipped with a steel impact cap. Ten impacts have been performed on each asperity, and for each, the response of all nine sensors has been monitored. \JS{All calibration experiments have been performed at room temperature.}

\begin{figure}[htb!]
\centering
\includegraphics[width=1\linewidth]{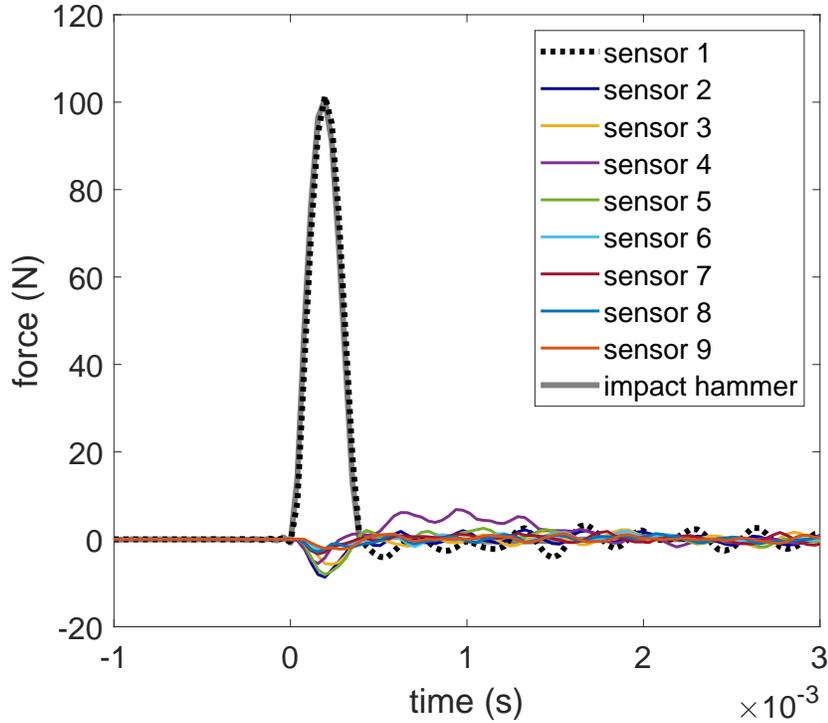}
\caption{Typical response of all nine piezoelectric sensors when asperity 1 is impacted with a hammer. Grey: force in hammer vs time. Black dashed (other colors, solid): concurrent force in piezoelectric sensor 1 (in the other eight sensors).}
\label{CurveHammer}
\end{figure}

Figure~\ref{CurveHammer} shows the results of a typical experiment. The impact duration is of order 0.5\,ms, with a maximum force of order 100\,N. The response of the piezoelectric sensor located just below the impacted surface asperity is essentially identical to that of the hammer. As a consequence, for each impact, we could determine a single scalar coefficient relating the force amplitude of the hammer’s response to the voltage amplitude from the stimulated piezoelectric sensor.
For each sensor, the sensitivity value was then taken as the average of those coefficients over the ten impacts on it. Note that those coefficients have been applied to all signals in Fig.~\ref{CurveHammer}.

When an impact is made above one of the nine piezoelectric sensors, the eight other sensors have a non-vanishing oscillating response, as seen in Fig.~\ref{CurveHammer}. This is an undesired feature of our instrumented slider, due to residual coupling between sensors. Those mechanical couplings are presumably due to the finite stiffness of the top plate of the slider, through which normal vibrations can propagate and affect all sensors\footnote{This assumption is substantiated by a test, performed on another (although very similar) version of the instrumented slider. The top and spring plates were removed before hammer impacts were applied directly on the spherical caps (label 3 in Fig.~\ref{fig:exploded}) glued on the piezoelectric sensors. In those conditions, the oscillating coda of the signals were significantly reduced.}. This interpretation is consistent with the results of the modal analysis presented in section~\ref{sec:fem}, where it is seen that the first eigenmodes of the slider are related to deformations of the top plate involving  its whole surface, and not localised around a single sensor. In addition, the first predicted eigenmodes have frequencies expected between 1.5 and 2\,kHz, which can easily be excited by the considered impacts of duration of the order of, but shorter than, 1ms. And indeed, in the various coda of Fig.~\ref{CurveHammer}, one can see oscillations with a typical frequency of order a few\,kHz (see, e.g., the coda of sensor 1 at about 3\,kHz).

\begin{figure}[htb!]
\centering
\includegraphics[width=0.9\textwidth]{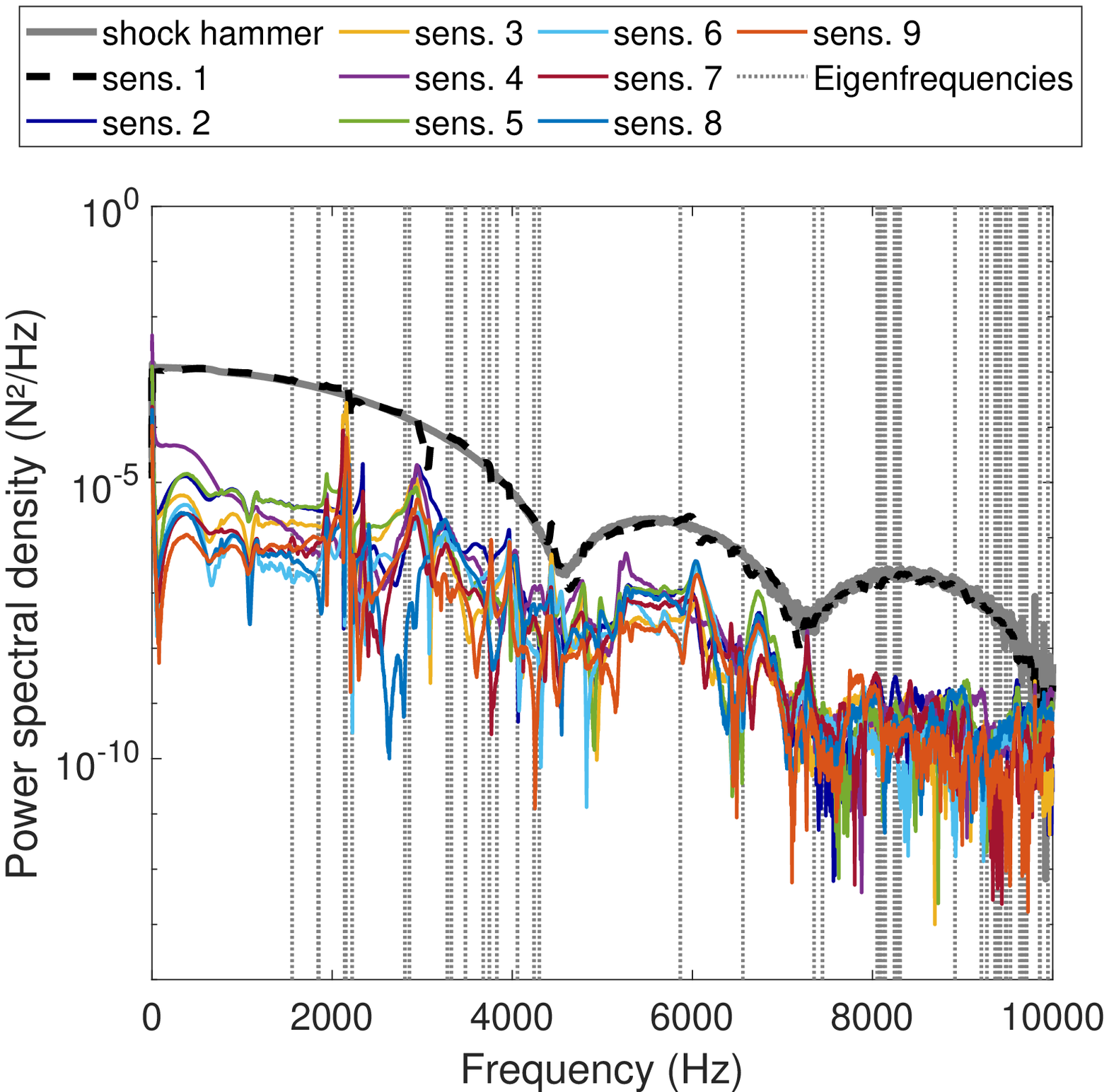}
\caption{Average PSDs of the nine output signals from the piezoelectric sensors and from the impact hammer, over ten impacts on asperity 1, one of which is shown in Fig.~\ref{CurveHammer}. Dotted vertical lines: eigenfrequencies predicted by the FE modal analysis of section \ref{sec:fem}.}
\label{fft}
\end{figure}

The frequency contents of the signals illustrated in Fig.~\ref{CurveHammer} can be seen in Fig.~\ref{fft}, which shows the average Power Spectral Density (PSD) over the ten impacts  performed on asperity 1. It appears that the spectra of the hammer and of the directly excited sensor are equal in the whole frequency range up to 10\,kHz, meaning that mechanical energy is actually injected and accurately measured by the piezoelectric sensor up to 10\,kHz. Note that the shape of the PSD, with its local minima at about 4700 and 7300\,Hz, is fully consistent with that of a half-sine wave of duration 0.33\,ms, which is a good approximation of the force signal in the impact hammer. As expected, for most of the frequencies in the range [1--1000]\,Hz, the other spectra are much weaker, by more than one order of magnitude.
However, the spectra of the sensors that are not directly excited by the hammer exhibit a series of peaks, whose amplitudes almost reach that of the excited sensor. Those peaks are interpreted as the eigenmodes of the slider, which is substantiated by the good matching between the frequency bands in which the peaks in the PSD are found, and those in which the eigenmodes have been predicted using the modal analysis of section~\ref{sec:fem} (Figs.~\ref{fig:Harmonic} and~\ref{fft}). Note that Fig.~\ref{fft} is actually in nice qualitative agreement with the FE predictions of Fig.~\ref{fig:Harmonic}: sensors 2 and 4 are also among the ones having the strongest response when asperity 1 is excited, and the signals of the non-excited sensors is enhanced in the same frequency bands (especially visible around 2 and 3\,kHz).
Overall, we emphasise that the large difference in amplitude between the excited asperity and the others, for all frequencies below 2\,kHz, is a significant success, because it opens to possibility to identify the impacted asperity from the unsollicited ones.

\subsection{Impact localisation and force estimation} \label{sec:localization}
Based on the experimental results of section~\ref{sec:calibration}, we now discuss the expected capabilities of our instrumented slider to locate in space and time, and to measure the amplitude of, a single impact among a random series of impacts. Given the large altitude of the nine spherical surface asperities with respect to the surface plate (about 3\,mm), tracks free from high-aspect-ratio protuberances will make contact with the slider only at those surface asperities, and not via the plate. Spatially locating an impact thus reduces to identifying which surface asperity is involved among the nine possible ones.

\subsubsection*{Model of the sensors' outputs}
Assuming that the slider behaves linearly \JS{and that, consequently, the principle of superposition holds}, the time output of the $j^{\text{th}}$ piezoelectric sensor, $S_j(t)$, can thus be written as \JS{a sum over the contributions of the individual forces on all nine surface asperities, as follows~\cite{roddier_distributions_1991,lebot_foundation_2015}}:
\begin{equation}
    S_j(t) = \sum_{i = 1} ^ 9 {\int_0 ^ \infty {C_{ji} (\tau) F_i (t-\tau) \mathrm{d}\tau}},\label{eq:linear}
\end{equation}
where $F_i(t)$ is the force signal applied on the $i^{\text{th}}$ surface asperity, and $C_{ji}(t)$ is the response function of sensor $j$ to an infinitely short impact on asperity $i$. Estimating the precise form of the $9 \times 9=81$ functions $C_{ji}$ is a very challenging task that we do not attempt to perform here. Instead, we will make some assumptions enabling simplification of Eq.~(\ref{eq:linear}) in an analytical way, thus providing a direct understanding of its metrological implications.

We first neglect any output from the piezoelectric sensors when no force is applied to the slider, i.e. we neglect the codas in Fig.~\ref{CurveHammer}. During an impact, we already noted in section~\ref{sec:calibration} that the output of the impacted sensor faithfully represents the time evolution of the external force. What can be seen from Fig.~\ref{CurveHammer} is that, for all the other eight sensors, the output during the impact has a bell shape with a small, negative amplitude. Although each of those signals has an individual extremum at a different instant, we now further assume that they all have the same shape as the external force (but different amplitudes). The two above-mentioned assumptions allow us to reduce the response functions to:
\begin{equation}
    C_{ji} (\tau) = a_{ji} \delta (\tau), \label{eq:dirac}
\end{equation}
where the coefficients $a_{ji}$ are assumed to be constants (unit V/N), and $\delta$ is the Dirac delta function. The coefficients $a_{jj}$ correspond to the individual sensitivities of the piezoelectric sensors, whose (positive) value has been estimated as described in section~\ref{sec:calibration} and used to plot sensor outputs in Newtons in Fig.~\ref{CurveHammer}. In a similar way, we evaluated the values of the $a_{ji, i \neq j}$ as the average value, over ten stimuli by the impact hammer on surface asperity $j$, of the extremum value of the output of sensor $i$ over the duration of the impact. If the instrumented slider was perfect, we would have $a_{ji} = 0$ for all $i \neq j$. In contrast, as seen in Fig.~\ref{CurveHammer} from the signals of sensors 2 to 8 during the impact, the $a_{ji, i \neq j}$ have non-vanishing (negative) values, which characterise cross-talk between the outputs from different sensors.

Inserting Eq.~(\ref{eq:dirac}) into Eq.~(\ref{eq:linear}), and dividing by $a_{jj}$, we obtain the following simplified relationship:
\begin{equation}
   \frac{S_j (t)}{a_{jj}} = F_j (t) + \sum_{i=1,i \neq j} ^ 9 \frac{a_{ji}}{a_{jj}} F_i (t).\label{eq:simple}
\end{equation}
The quantity $S_j (t)/a_{jj}$ is an estimate of the impact force, $F_j(t),$ based on the sensor output, $S_j(t)$. If all $a_{ji, i \neq j}=0$, the estimate is accurate because $S_j (t)/a_{jj}=F_j(t)$. In reality, the estimate is biased by an error $E(t)=\sum_{i=1,i \neq j} ^ 9 p_{ji} F_i (t)$, which depends both on the amplitudes of the external forces, $F_i(t)$, and on the ratios of cross-talk over sensitivity coefficients,
$p_{ji}=a_{ji}/a_{jj}$. By definition, all diagonal elements of the $p_{ji}$ matrix are 100\%. Except for sensor 4,
all non-diagonal terms remain smaller than 10\% (Fig.~S4), indicating a relatively small cross-talk between outputs.

\subsubsection*{Localisation strategies}
Let us first consider the case where the slider is submitted to a series of impacts happening sequentially, without any overlap in time: for each external force signal $F_j(t)$ on asperity $j$, $F_{i,i \neq j}=0$ in the same time interval. We are in the favourable case, already tested with the impact hammer, where signals consist of a succession of single events like the one shown in Fig.~\ref{CurveHammer}. For each, the asperity involved will be the one giving the largest signal (say $j$); its amplitude will be accurately estimated by $S_j (t)/a_{jj}$, while the other signals (for $i \neq j$) can be overlooked.

The situation is more complex when several impacts happening on different asperities overlap in time. In section~\ref{sec:resultsFlat}, we will argue theoretically and confirm experimentally that the number of asperities in simultaneous contact with a rigid track is not expected to exceed three. For the rest of this section, we will thus consider the unfavourable case of three simultaneous impacts, all with a force evolution proportional to the same function $F_0(t)$. Without loss of generality, we will assume that those three asperities are labelled 1 to 3, and that $F_1(t)=F_0(t)$, $F_2(t)=\alpha F_0(t)$, $F_3(t)=\beta F_0(t)$, with $1>\alpha>\beta$. Finally assume that all $p_{ij}$ are such that $\left|p_{ji}\right| \le p$. In those conditions, Eq.~(\ref{eq:simple}) can be recast into:
\begin{align}
S_1/a_{11} &= F_0(t) \left(1 \pm p(\alpha + \beta)\right),\label{eq:simimp1}\\
S_2/a_{22} &= \alpha F_0(t) \left(1 \pm p(1 + \beta)/\alpha \right),\label{eq:simimp2}\\
S_3/a_{33} &= \beta F_0(t) \left(1 \pm p(1+ \alpha)/\beta\right),\label{eq:simimp3}\\
S_j/a_{jj} &= \pm F_0(t) p(1 + \alpha + \beta), \quad j>3.\label{eq:simimpj}
\end{align}

Those equations indicate a straightforward data analysis when three simultaneous impacts are suspected. First identify the three sensor outputs giving the largest impact forces and record the three corresponding peak values. The largest gives $F_0$, while the ratio of the other two over $F_0$ give $\alpha$ and $\beta$. With $p=0.1$ being a representative value for our slider (Fig.~S4), Eq.~(\ref{eq:simimpj}) indicates that 0.3$F_0$ is a conservative estimate ($p(1+\alpha+\beta)<0.3$) of the experimental noise. In other words, all peaks larger than this value can be safely considered to be true impacts and not the results of some cross-talking between outputs. All peaks below may be discarded. The amplitude of the impact force on the successfully localised asperities can thus be conservatively estimated, using Eqs.~(\ref{eq:simimp1}),~(\ref{eq:simimp2}) and~(\ref{eq:simimp3}), to be $F_0 \pm 20\%$, $\alpha F_0 \pm \frac{20}{\alpha}\%$ and $\beta F_0 \pm \frac{20}{\beta}\%$, respectively.

\section{Experimental results}\label{sec:ExpResults}
\begin{figure}[htb!]
\centering
\includegraphics[width=0.9\textwidth]{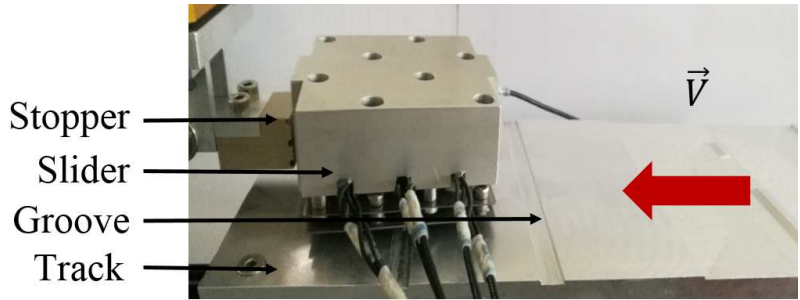}
\caption{Picture of the mechanical setup. A textured metallic track is translated horizontally at constant velocity, $V$, while the slider is held fixed in the laboratory frame by a stopper.}
\label{fig:ExpPicture}
\end{figure}

In order to illustrate the tribological relevance of the instrumented slider described and characterised in section~\ref{sec:slider}, we now slide it on various model topographies. A stainless steel track has been textured with macroscopic obstacles, grooves and bumps of sufficient amplitude and lateral size to induce controlled changes in the distribution of spherical asperities in contact during sliding. Figure~\ref{fig:ExpPicture} shows a picture of the instrumented slider, laid on the textured track (upside down compared to Fig.~\ref{Pad}). The latter is attached on a motorised (Kollmorgen, AKM Series servo-motor) linear translation stage (Misumi LX26). When the track is translated, the displacement of the slider is prevented by a stopper, fixed in the laboratory frame, and equipped with a horizontal steel cylinder (1\,mm diameter) which comes into contact with the vertical plane of the lateral arm of the slider's base which is close to asperity 4 (see Fig.~\ref{Pad}). The tangential force exerted by the stopper is measured by a piezoelectric sensor (Kistler 9217A, with charge amplifier Kistler 5015A).

The principle of the experiments is the following. The normal load on the interface, approximately $P=2.5$\,N, is the weight of the instrumented slider. The textured track is driven horizontally at a constant velocity, $V$, in the range [0.5--5]\,mm/s. Two different illustration cases will now be analysed: when the slider explores a nominally flat part of the track (section~\ref{sec:resultsFlat}) and when one of the spherical asperities passes above a macroscopic hole (section~\ref{sec:resultsDefects}). \JS{All sliding experiments have been performed at room temperature.} \JS{Note that the contact size on each surface asperity of the slider is expected to be smaller than 36\,$\mu$m, as estimated using Hertz's theory with a conservative value of the normal force of 2.5\,N, radius of curvature $R$=0.75\,mm, Young's modulus $E$=210\,GPa and Poisson's ratio $\nu$=0.3 for the asperity. As long as the characteristic distance between two neighbouring topographical protuberances of the track remains larger than those 36\,$\mu$m, the measured contact forces can safely be interpreted as that of a single asperity/asperity contact. As will be seen, this is the case for all experiments in section~\ref{sec:ExpResults}. Also note that inspection of the slider's surface asperities at the end of the experiments has not revealed any significant sign of wear, as expected for the low severity conditions of our sliding experiments (small normal load and relatively small sliding velocities).} 

\begin{figure*}[hbt!]
\centering
\includegraphics[width=0.99\textwidth]{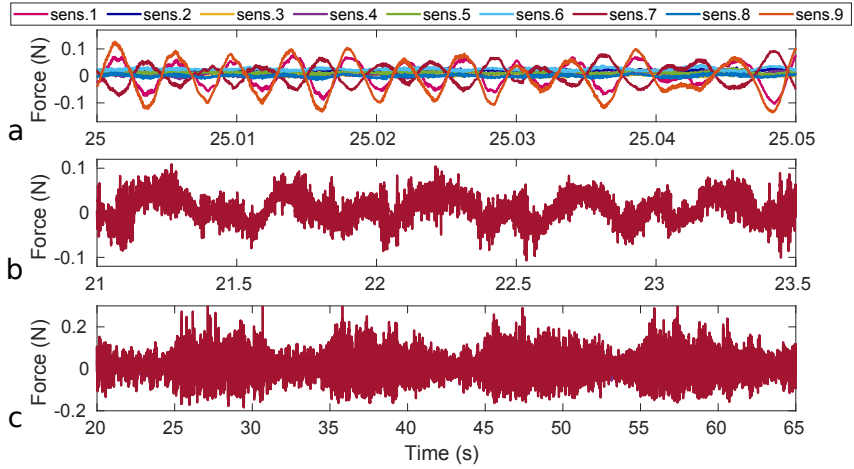}
\caption{Typical time evolution of the asperity forces recorded during steady sliding on a nominally flat part of the track at $V=0.5$\,mm/s, for various time-windows: 50\,ms (a, all asperities), 2.5\,s (b, asperity 7) and 45\,s (c, asperity 7).}
\label{fig:Machine9Piezo200Hz}
\end{figure*}

\subsection{Sliding on a nominally flat surface}\label{sec:resultsFlat}
Figure~\ref{fig:Machine9Piezo200Hz} shows a typical recording of the nine piezoelectric sensors during steady sliding over a nominally flat part of the textured track, at a velocity $V=0.5$\,mm/s. Surprisingly, instead of the expected constant force signals at different values, the sensors' outputs all have a vanishing mean value, and reveal a variety of additional features, at different timescales (different panels of Fig.~\ref{fig:Machine9Piezo200Hz}). While the zero mean value originates from the fact that frequencies smaller than about 1\,Hz are filtered out by the measurement chain (already described in section~\ref{sec:instrum} and further discussed in section~\ref{sec:resultsDefects}), the scope of this section is to unravel the tribological origin of those additional features.

Over a short time window of 50\,ms (Fig.~\ref{fig:Machine9Piezo200Hz}(a)), two main observations can be made: (i) three sensors (sensors 1, 7 and 9) produce a signal with a significantly larger amplitude than the other six and (ii) those three signals have a clear periodicity. The first observation suggests that only three surface asperities are in actual contact with the track during sliding. The fact that not all the slider's asperities can touch a rigid plane at the same time is a natural consequence of having stiff asperities with altitudes distributed around a mean plane. As can be seen in Table~S1, the altitudes of the two highest asperities differ by 13\,$\mu$m. An estimate of the order of magnitude of the normal force on the highest asperity necessary to deform it by such an amount (and thus, assuming that the slider cannot rotate, to allow both asperities to be in contact simultaneously) can be obtained using \JS{the same} Hertz's theory \JS{as in the preamble of section~\ref{sec:ExpResults}. We find that} such a normal force is about 400\,N, i.e. way larger than the slider's weight ($P$=2.5\,N). Thus, in the light normal load conditions explored here, only three asperities are expected to touch the track, to satisfy isostatic equilibrium, which explains our first observation.

In order to predict which are the three  asperities in contact among the nine potential asperities, a simple multi-asperity contact model has been used, fully described in section~S2 (in SM). The model is dynamic, but when constant stimuli are imposed, it provides results relevant to quasi-static configurations. In particular, when the model is used with a vanishingly small sliding velocity, it predicts, based on the experimental in-plane positions and altitudes of all nine spherical asperities (Tab.~S1), which are the three that are in contact with a given track topography, $h(x,y)$. In the present case in which $h$ can be assumed to be homogeneous, the model predicts that asperities 1, 7 and 9 are in contact to satisfy isostatic equilibrium. Those asperities are the same as the ones that are active in Fig.~\ref{fig:Machine9Piezo200Hz}(a). The model in section~S2 thus fully explains both the number and identities of the asperities in contact. 

\begin{figure}[htb!]
\centering
\includegraphics[width=0.8\linewidth]{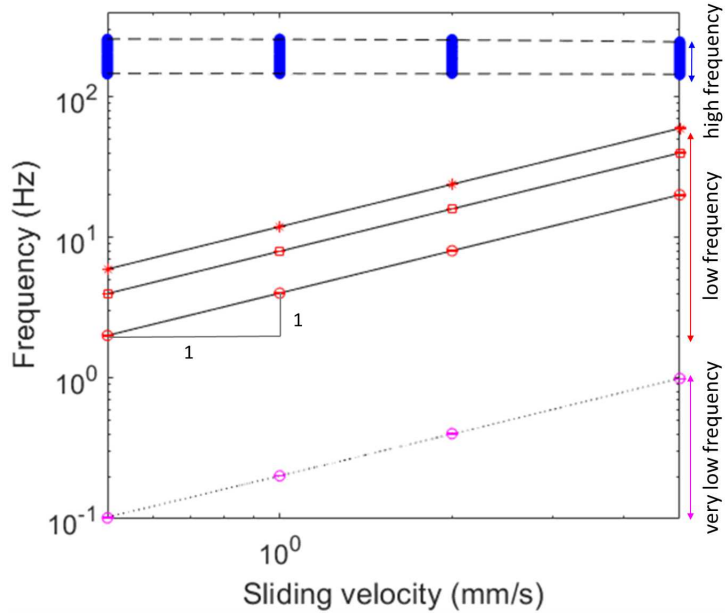}
\caption{Evolution of the various frequencies identified in the force signals, as a function of sliding velocity. Pink: very low frequencies seen in Fig.~\ref{fig:Machine9Piezo200Hz}(c). Red: low frequencies seen in Fig.~\ref{fig:Machine9Piezo200Hz}(b). Circles: fundamental. Squares (+): first (second) harmonic. The high-frequency band ([150--250]\,Hz) responsible for the oscillations seen in Fig.~\ref{fig:Machine9Piezo200Hz}(a) is shown in blue and delineated with dashed lines. Solid and dotted lines: linear fits.}
\label{fig:MachineFreqvsV}
\end{figure}

The second observation in Fig.~\ref{fig:Machine9Piezo200Hz}(a) is the periodicity of the local force signals, at a frequency about 200\,Hz. Inspection of the PSDs of all force signals for all sliding velocities revealed a common frequency band [150--250]\,Hz with a significantly larger amplitude, responsible for the oscillations in Fig.~\ref{fig:Machine9Piezo200Hz}(a). The sliding-velocity-independence of this high-frequency band (Fig.~\ref{fig:MachineFreqvsV}) suggests that the oscillation is due to eigenmodes of the experimental system. Since all eigenmodes of the clamped slider were above 1\,kHz (section~\ref{sec:calibration}), we presume that the 200\,Hz mode is associated with the vibration of the bundles of cables (see Fig.~\ref{fig:ExpPicture}), which is now possible with the slider freely standing on the track. 

\begin{figure}[b!]
\centering
\includegraphics[width=0.99\columnwidth]{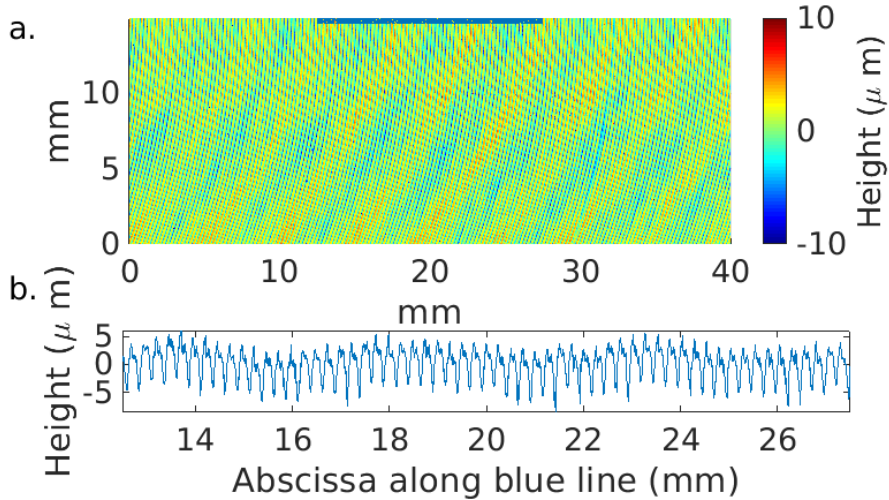}
\caption{(a): 2D map of the topography of the nominally flat sliding surface, as measured from interferometric profilometry. (b): profile extracted along the segment drawn in the top panel, which is locally orthogonal to the grooves.}
\label{fig:MachineTopo}
\end{figure}

Another periodicity can be found in the force signals of the three contacting asperities, at a larger time scale, as illustrated in Fig.~\ref{fig:Machine9Piezo200Hz}(b). For $V=0.5$\,mm/s, the observed frequency is about 2\,Hz. Such a low frequency suggests that the nominally flat track surface is actually decorated with a periodic topography. This is indeed confirmed in Fig.~\ref{fig:MachineTopo}(a), which shows the topography of a portion of the track surface. It is made of characteristic trochoidal grooves originating from the face milling process used to prepare the surface. As seen in Fig.~\ref{fig:MachineTopo}(b), those grooves have a period of 248$\pm$12\,$\mu$m (as measured from the corresponding peak in the Fourier transform of the profile in Fig.~\ref{fig:MachineTopo}(b))\footnote{\JS{This period is much larger than the 36\,$\mu$m estimated for the maximum contact size of a surface asperity of the slider (see preamble of section~\ref{sec:ExpResults}).}} and typical peak-to-peak amplitude 8\,$\mu$m. Assuming that the profile in Fig.~\ref{fig:MachineTopo}(b) would be a perfect sine wave, the radius of curvature of the bottom parts of those grooves would be about 400\,$\mu$m, i.e. smaller than that of the spherical asperities (750\,$\mu$m).
Thus, the slider's asperities cannot explore the whole topography, but only the crests of the grooves. More quantitatively, we calculated that a sphere of radius 750\,$\mu$m cannot go deeper than 6.5\,$\mu$m into the above-mentioned sine wave without yielding an unphysical interpenetration of the two solids.
We have thus applied the model of section~S2 to static topographies in which the track altitude at the location of asperities 1, 7 and 9 were offset by either 0 or -6.5\,$\mu$m. We tested all eight combinations of those two offsets at the location of the three asperities, as estimators of the strongest possible differences with respect to a perfectly flat track. Only two combinations led to a change in the predicted asperities in contact: asperity 4 takes over asperity 7 when both asperities 1 and 7 are simultaneously above the deepest point of a groove. Given the small probability of such a situation, it is reasonable to consider it as irrelevant in practice. 
Consequently, the amplitude of the topography of the milling-induced grooves is not sufficient to induce detectable modifications of the asperities in contact with respect to a perfectly smooth track, explaining why the three asperities in contact (1, 7 and 9) remain the same all along the sliding experiment on the nominally flat part of the track.

To ascertain the topographical origin of the force oscillations seen in Fig.~\ref{fig:Machine9Piezo200Hz}(b), we now compare the frequency of those oscillations with that expected from sliding on the grooves. We performed sliding experiments at different velocities, $V$, and for each we calculated the PSD. In the frequency range [1--100]\,Hz, they all contain a well-defined main peak (fundamental) and a series of harmonics. The latter are due to the non-harmonic shape of the periodic pattern. The frequencies of the fundamental and of its two first harmonics are shown as a function of $V$ in Fig.~\ref{fig:MachineFreqvsV}. All three frequencies are found proportional to $V$, as demonstrated by the slope 1 in the log-log representation of Fig.~\ref{fig:MachineFreqvsV}. This shows that the temporal periodicity actually originates from a spatial periodicity. For the fundamental, the coefficient of proportionality, which corresponds to the spatial period, is found to be $249\pm$2\,$\mu$m. This value is in good agreement with the period of the grooves found by profilometry (248$\pm$12\,$\mu$m).

A third periodicity can be seen in Fig.~\ref{fig:Machine9Piezo200Hz}(c), as a very low frequency modulation of the signal's envelope. As seen in Fig.~\ref{fig:MachineFreqvsV}, the frequency of this modulation is also proportional to the sliding velocity, thus indicating a topographical origin for it. The corresponding wavelength is about 5.0\,mm, a length scale that can indeed be distinguished as a waviness in the topographical signal of Fig.~\ref{fig:MachineTopo}. Such a waviness is classically attributed to an undesired coaxiality error between the centerlines of the spindle and the milling tool.

Overall, with the above analyses, we took advantage of undesired residual topographical features on the flat parts of the track to show the capabilities of our instrumented slider. Not only is it able to detect those residual, micrometric features at different time/length scales, but, as desired, it also quantifies the elicited force fluctuations, and indicates which asperities are excited.

\begin{figure*}[hbt!]
\centering
\includegraphics[width=0.99\textwidth]{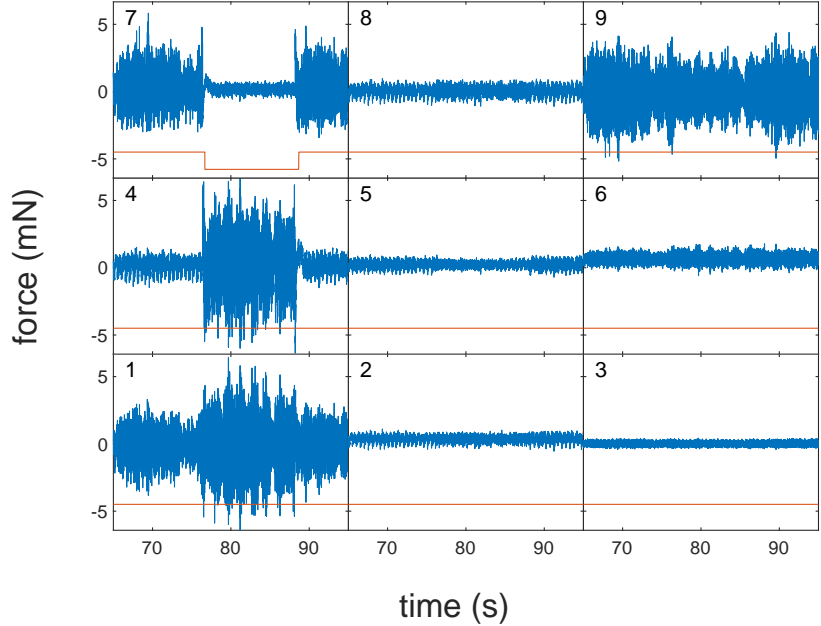}
\caption{Blue: Force on the nine asperities (asperity label in top left corner of each panel) when asperity 7 passes over a deep, 6\,mm long groove at $V=0.5$\,mm/s. Red: sketch of the macroscopic topography of the track (starting at 0) seen by each asperity.}
\label{fig:defect7_500}
\end{figure*}

\subsection{Sliding on a macroscopic topographic defect}\label{sec:resultsDefects}
In section~\ref{sec:resultsFlat}, we have seen that, when the track is nominally flat, asperities 1, 7 and 9 are in contact at all times. We will now consider a situation in which the asperities in contact must change during sliding, because one of those three asperities passes over a hole deep enough to prevent any possible contact for a certain sliding distance. Such a situation is illustrated in Fig.~\ref{fig:defect7_500}, which shows the recorded force signals on all nine asperities over a time window bracketing the passage of asperity 7 over a 6\,mm long hole.

Before reaching the hole, the situation is the same as in section~\ref{sec:resultsFlat}: asperities 1, 7 and 9 are in contact, as indicated by a significantly larger signal width than the other six, due to the high frequency oscillations seen in Fig.~\ref{fig:Machine9Piezo200Hz}(a). When asperity 7 passes over the hole, its force signal fluctuations suddenly die out, indicating that, as expected, it has lost contact with the track. Simultaneously, the force fluctuations on asperity 4 abruptly increase, showing that asperity 4 is now sliding on the track. Application of the model of section~S2 to the new topography seen by the slider consistently predicts that asperities 1, 4 and 9 is the triplet satisfying the new isostatic equilibrium, thus explaining the switching of activity between asperities 7 and 4. When asperity 7 reaches the end of the hole, the slider and hence all force signals recover their initial configuration.

\begin{figure}[htb!]
\centering
\includegraphics[width=0.9\columnwidth]{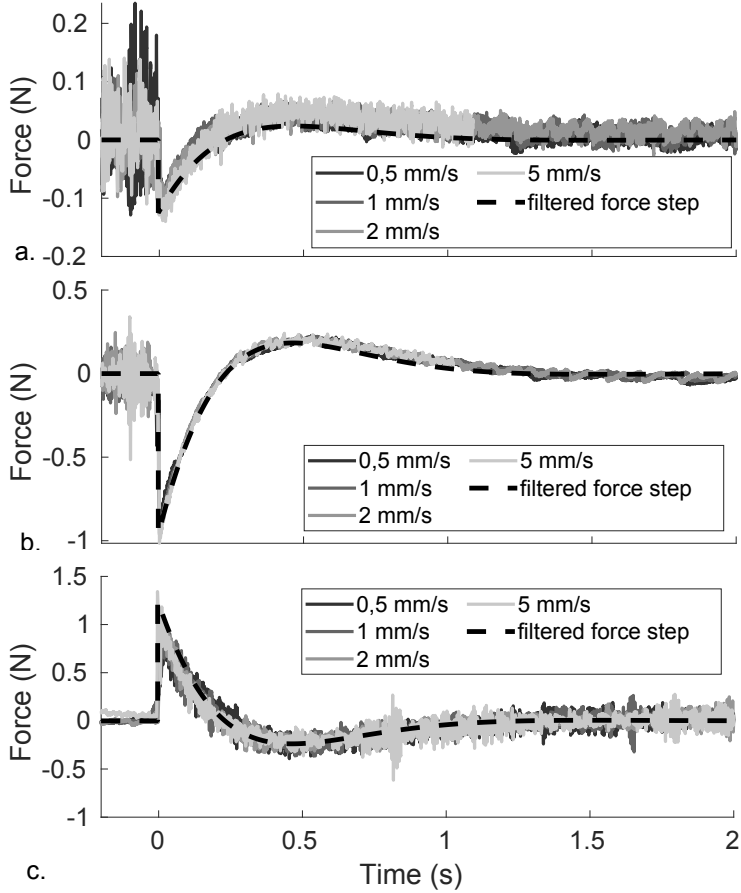}
\caption{Force transients on asperity 7 (a) or 9 (b and c) when it suddenly looses (a and b) or gets back into contact (c) with the track for various sliding velocities. Dashed lines: instantaneous negative (a and b) or positive (c) force drops filtered by the second-order high-pass filter of Eq.~(\ref{eq:filter}).}
\label{fig:defect7_zoom_allV}
\end{figure}

We emphasise that, in Fig.~\ref{fig:defect7_500}, the fact that a given asperity is in contact can only be seen from the amplitude of the force fluctuations discussed in section~\ref{sec:resultsFlat}. Contact/non-contact states do not translate into a finite/vanishing average level of the signals, because of the high-pass filtering effect of the conditioning amplifiers used on each sensor's output. Those amplifiers are claimed by the manufacturer to behave like a 1\,Hz first order high-pass filter, meaning that, after about 1\,s, any constant force applied on an asperity will be filtered out from the outputs, yielding force signals centered on zero. Nevertheless, the force signal on asperity 7 exhibits transients at the edges of the hole, when asperity 7 gets out of (or back into) contact with the track, which are signatures of the corresponding impact-like changes in average normal force on the asperity. Figure~\ref{fig:defect7_zoom_allV}(a) overplots the transients associated with the force drop at the entry of the hole, for all tested sliding velocities. As can clearly be seen, they all have the same amplitude and shape, with the same characteristic timescale for the force relaxation. The very same phenomenology is observed in other experiments in which other asperities pass over the hole, as illustrated in Fig.~\ref{fig:defect7_zoom_allV}(b) on the case of asperity 9.

The time evolution of all signals in Fig.~\ref{fig:defect7_zoom_allV}(a) and (b) can be nicely captured as the response of \JS{a} second-order high-pass filter \JS{with the following transfer function vs frequency $f$,}
\begin{equation}
  HP(f)=\frac{-f^2/f_c^2} {1+j f/(Q f_c)-f^2/f_c^2},
  \label{eq:filter}
\end{equation}
\JS{when} applied to a negative step-change (force drop) on the asperity. $Q=0.656$ is the quality factor, $f_c=0.74$\,Hz is the cutoff frequency \JS{and $j$ is the imaginary unit}. This cutoff frequency is close to the 1\,Hz expected for the filtering effect of the conditioning amplifiers. The fact that the filter is of second order rather than the expected first order is presumably due to a non-negligible effect of the rest of the measurement chain on the final outputs.

Importantly, the peak in the filtered response is equal to the amplitude of the force step-change, meaning that the amplitude of the peak of the transient is actually a measurement of the force drop undergone by the asperity when it start passing over the hole. In panel (a) of Fig.~\ref{fig:defect7_zoom_allV}, one can thus estimate the force drop on asperity 7 to be about -0.12\,N, while it is about -0.96\,N in panel (b) for asperity 9. Those values actually compare very well with the force drops predicted by the model of section~S2: -0.11\,N for asperity 7, and -1.15\,N for asperity 9.

Note that transients with the opposite sign are observed in the force signals when the underlying force step is opposite, for instance when asperity 9 suddenly gets back into contact on the other side of the hole (Fig.~\ref{fig:defect7_zoom_allV}(c)). The transients are still well-captured by the filter of Eq.~(\ref{eq:filter}) (shown only for $V$=5\,mm/s on Fig.~\ref{fig:defect7_zoom_allV}(c)). However, in contrast with when the asperities loose contact with the track, when they get back into contact, the amplitude of the transient exhibit a non-negligible sliding-velocity dependence. It would correspond to a positive force-step of amplitude 0.8, 0.9, 1.1 and 1.25\,N for $V$=0.5, 1, 2 and 5\,mm/s, respectively. Such a velocity dependence is compatible with more violent lateral impacts of asperity 9 with the end corner of the hole as the sliding velocity is larger. This effect is not expected at the entry of the hole, explaining why the same amplitude was found for all sliding velocities in Figs.~\ref{fig:defect7_zoom_allV}(a) and (b).

Overall, the good agreement between the sensors' outputs and the filtered expected force jumps indicates that our instrumented slider can efficiently be used to measure the impact forces on its asperities. The results of Fig.~\ref{fig:defect7_zoom_allV} show that it is the case for abrupt step-changes in the force between two constant values. For more realistic impacts characterised by a short contact duration, i.e. sufficiently shorter than the cutoff timescale of about 1\,s, the high-pass filter will deform only slightly the signals, so that the sensors outputs are expected to provide, directly, a faithful image of the force evolution. To verify this, we applied the filter of Eq.~(\ref{eq:filter}) onto half-sine model impacts with different durations between 10\,s and 0.1\,ms. We found that the alteration of the force evolution is negligible (less than 2\% in amplitude) for impacts shorter than about 0.01\,s. This result strongly suggests that, during sliding of two randomly rough metallic surfaces, for which the typical impact duration is of the order or less than the millisecond~\cite{dang_direct_2013}, our slider will provide accurate measurements of the individual impact forces. 

\section{Conclusion}
We have introduced a 6\,cm-sized instrumented slider based on an array of \JS{nine} piezoelectric sensors, \JS{each monitoring a single spherical surface asperity with a well-known radius (0.75\,mm) and altitude (72\,$\mu$m standard deviation among the nine asperities). The slider} is able to \JS{measure} the normal components of the individual contact forces on all \JS{surface} asperities, enabling both spatial localisation and \JS{quantification} of the amplitude of the impacts on the slider's surface. The slider behaves statically below about 1.5\,kHz, and rejects less than 5\% of the force on a given asperity onto the other sensors. \JS{To demonstrate the capabilities of the slider, we performed a series of sliding experiments under the slider's own weight (2.5\,N), at constant velocity in the range [0.5--5]\,mm/s, against a model rough track.} We \JS{first} showed how the slider can successfully be used to \JS{identify} the \JS{periodic} force variations on each individual surface asperity elicited by \JS{regularly spaced, micrometric grooves left on a nominally flat part of the track's surface by the milling process used to prepare it. We then showed how the slider can quantitatively measure the impact-like forces induced}  by large \JS{topographical} defects such that an asperity suddenly loses (or gets back into) contact with the track.

Our instrumented slider thus appears as a promising tool for a variety of future tribological studies involving realistic randomly rough surfaces. In particular, it should be useful to better characterise the impacts at the origin of the roughness noise, the source of friction-induced vibrations, or the space-time patterns of the real contact fluctuations during sliding.

\newpage

\renewcommand{\thefigure}{S\arabic{figure}}
\renewcommand{\thetable}{S\arabic{table}}
\renewcommand{\thesection}{S\arabic{section}}
\renewcommand{\theequation}{S\arabic{equation}}
\setcounter{figure}{0}
\setcounter{equation}{0}
\setcounter{section}{0}

\section{Supplementary Figures and Tables}

\begin{figure}[h!]
\centering
\includegraphics[width=0.99\textwidth]{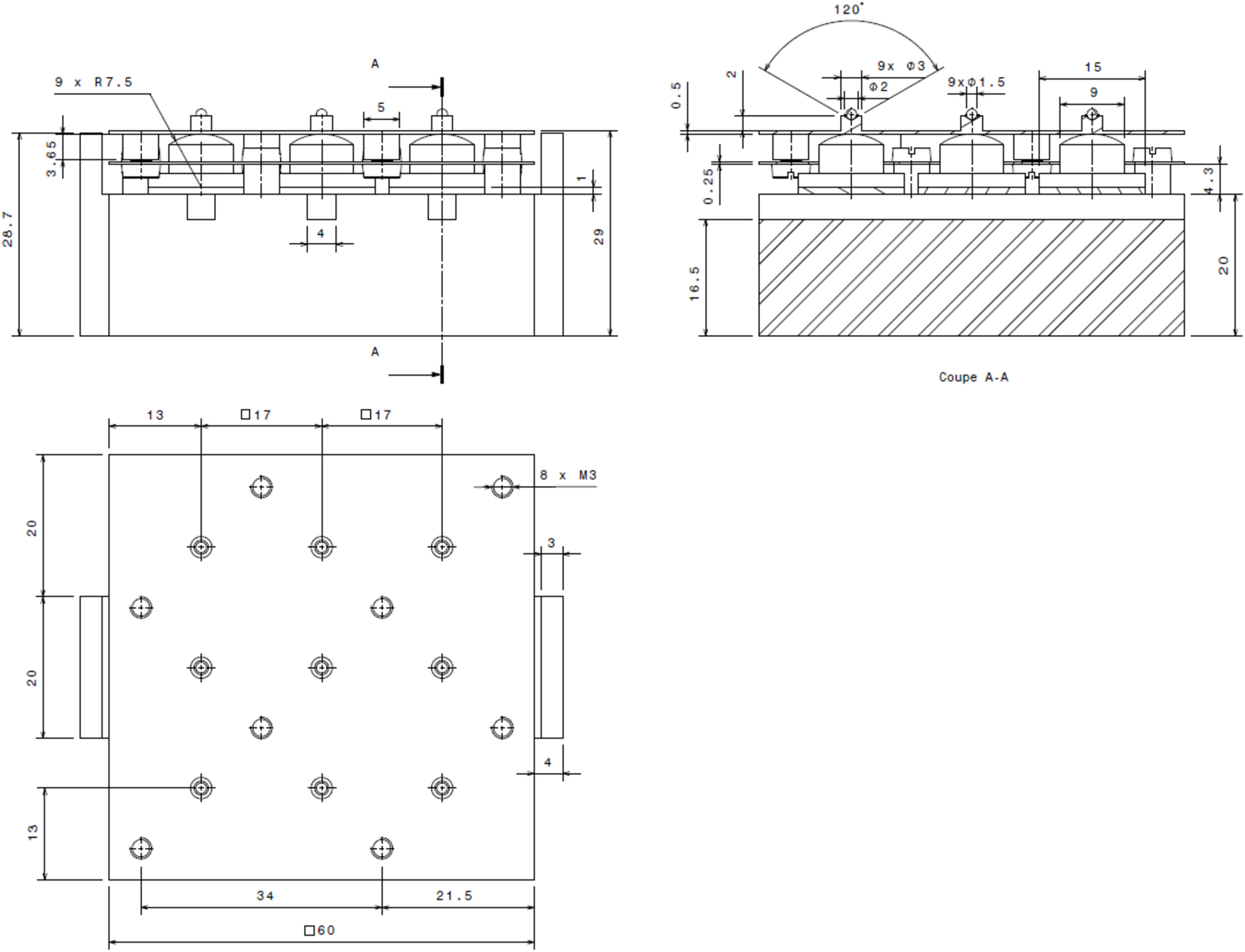}
\caption{Final design and dimensions of the slider (using the general tolerances ISO 2768mK).}
\label{Stat}
\end{figure}

\newpage

\begin{table}
\centering
\caption{Individuals coordinates of the model asperities labeled 1 to 9 in Fig.~3. $x$- and $y$-positions (in mm) are relative to the center of mass of the slider. The altitudes (in $\mu$m) are relative to the asperities' mean plane.}
\label{tab:altitudes}
\begin{tabular}{cccc}
\hline\noalign{\smallskip}
Asperity&Position $x$&Position $y$&Altitude\\
\noalign{\smallskip}\hline\noalign{\smallskip}
1 & 17.42 & 15.95 & 107\\
2 & 17.53 & -1.12 & -30\\
3 & 17.45 & -17.87 & -96\\
4 & 0.59 & 16.09 & 14\\
5 & 0.44 & -1.04 & -30\\
6 & 0.43 & -17.99 & 55\\
7 & -16.44 & 16.00 & -68\\
8 & -16.49 & 0.92 & -46\\
9 & -16.42 & -18.11 & 94\\
\noalign{\smallskip}
\end{tabular}
\end{table}\label{tab:altitudes}

\begin{table}
\centering
\caption{Materials properties used in the Finite Element model.}
\label{tab:materials}
\begin{tabular}{ccccc}
\hline\noalign{\smallskip}
Materials & Part label in Fig.~2 & Young Modulus $E$ ($GPa$) & Poisson's ratio $\nu$ & Density  $\rho$ ($kg/m^3$)\\
\noalign{\smallskip}\hline\noalign{\smallskip}
Aluminum & 1, 3 & 71 & 0.33 & 2770\\
Carbon steel & 4, 5, 6, 7, 8, 9 & 200 & 0.3 &7850\\
PZ 27 & 2 &$E_{xx}=E_{yy}=66.0$ $E_{zz}=84.3$ & 0.389 & 7700\\
\noalign{\smallskip}\hline
\end{tabular}
\end{table}

\newpage

\begin{figure}[h!]
\centering
\includegraphics[width=0.7\textwidth]{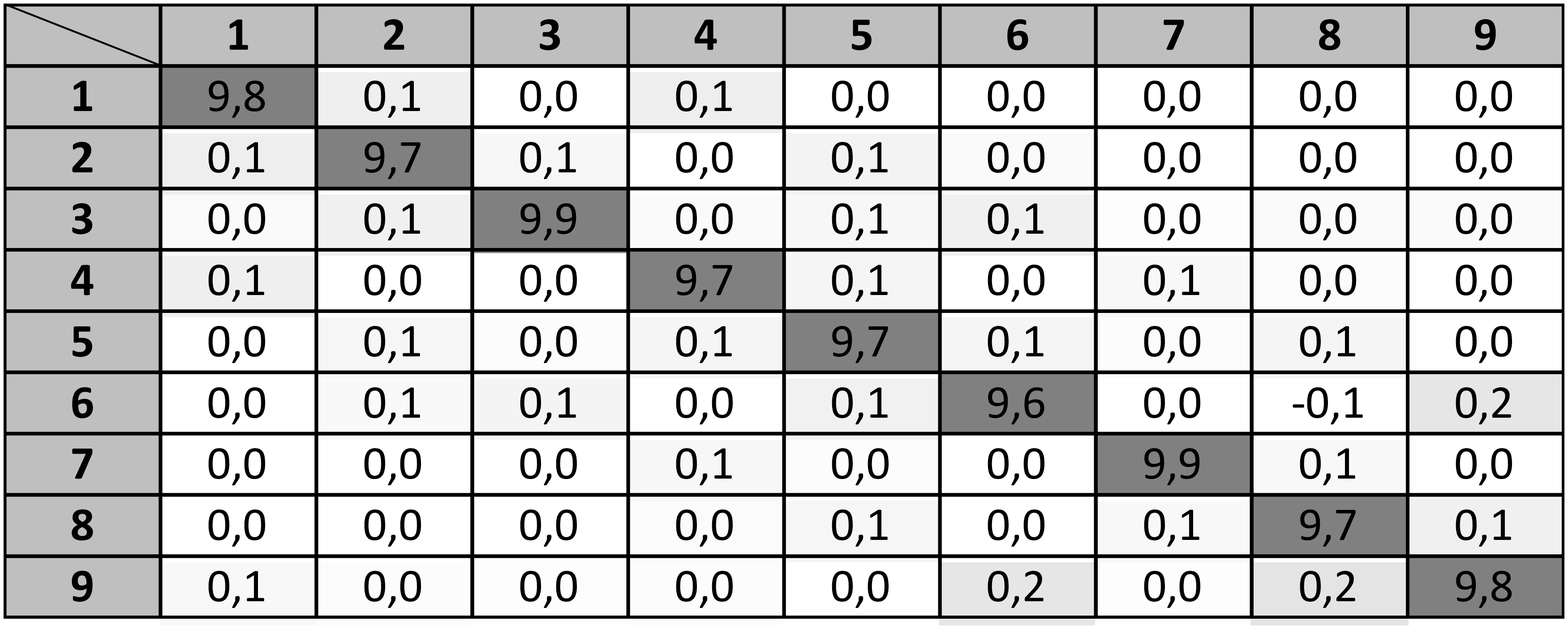}
\caption{Forces (in N) on the piezoelectric sensors when a static vertical force of 10\,N is applied to a single surface asperity, in our FE model. Row: label of the loaded asperity. Column: label of the sensor.}
\label{Stat}
\end{figure}

\newpage

\begin{figure}[h!]
\centering
\includegraphics[width=0.5\columnwidth]{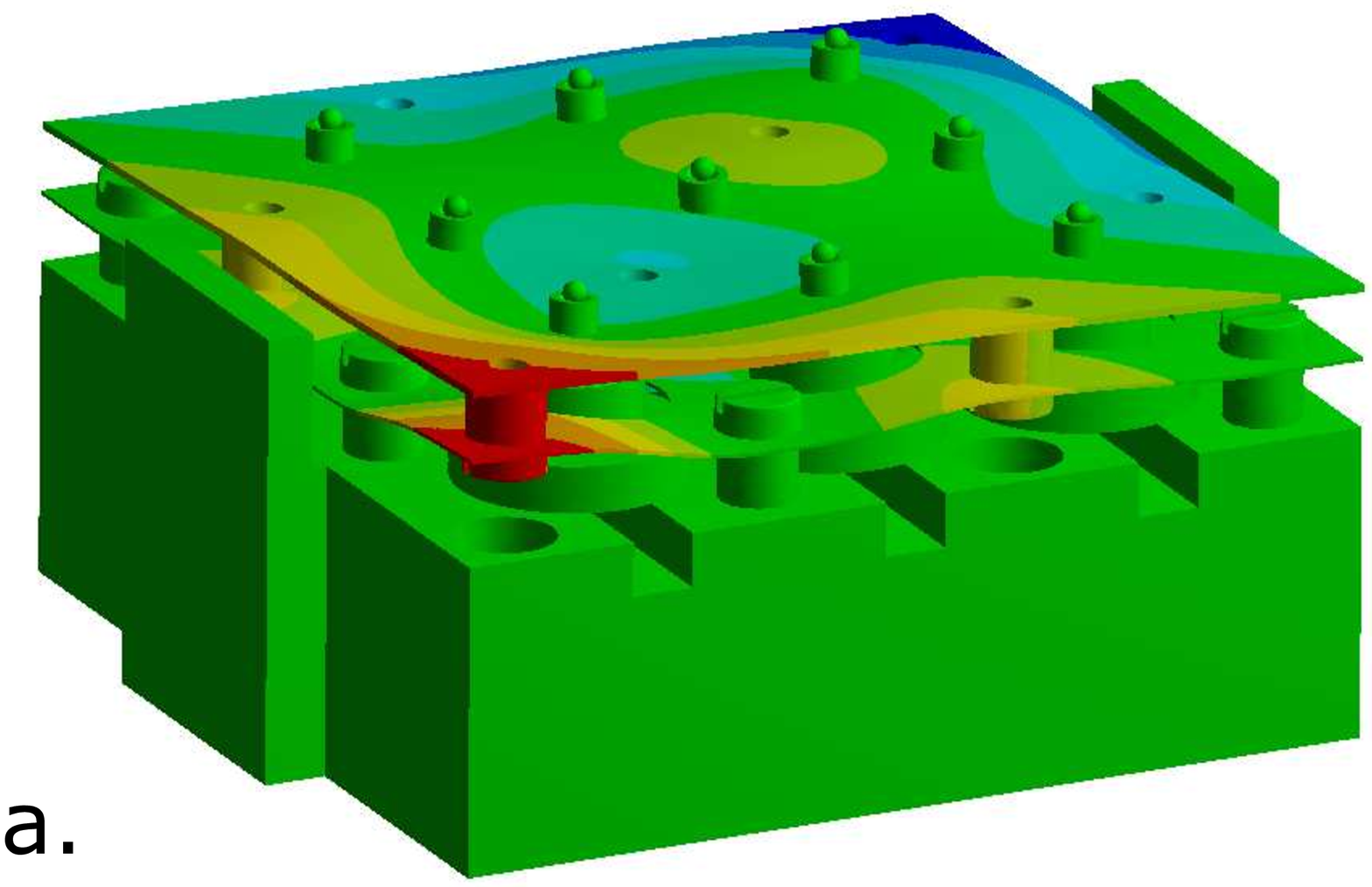}
\includegraphics[width=0.5\columnwidth]{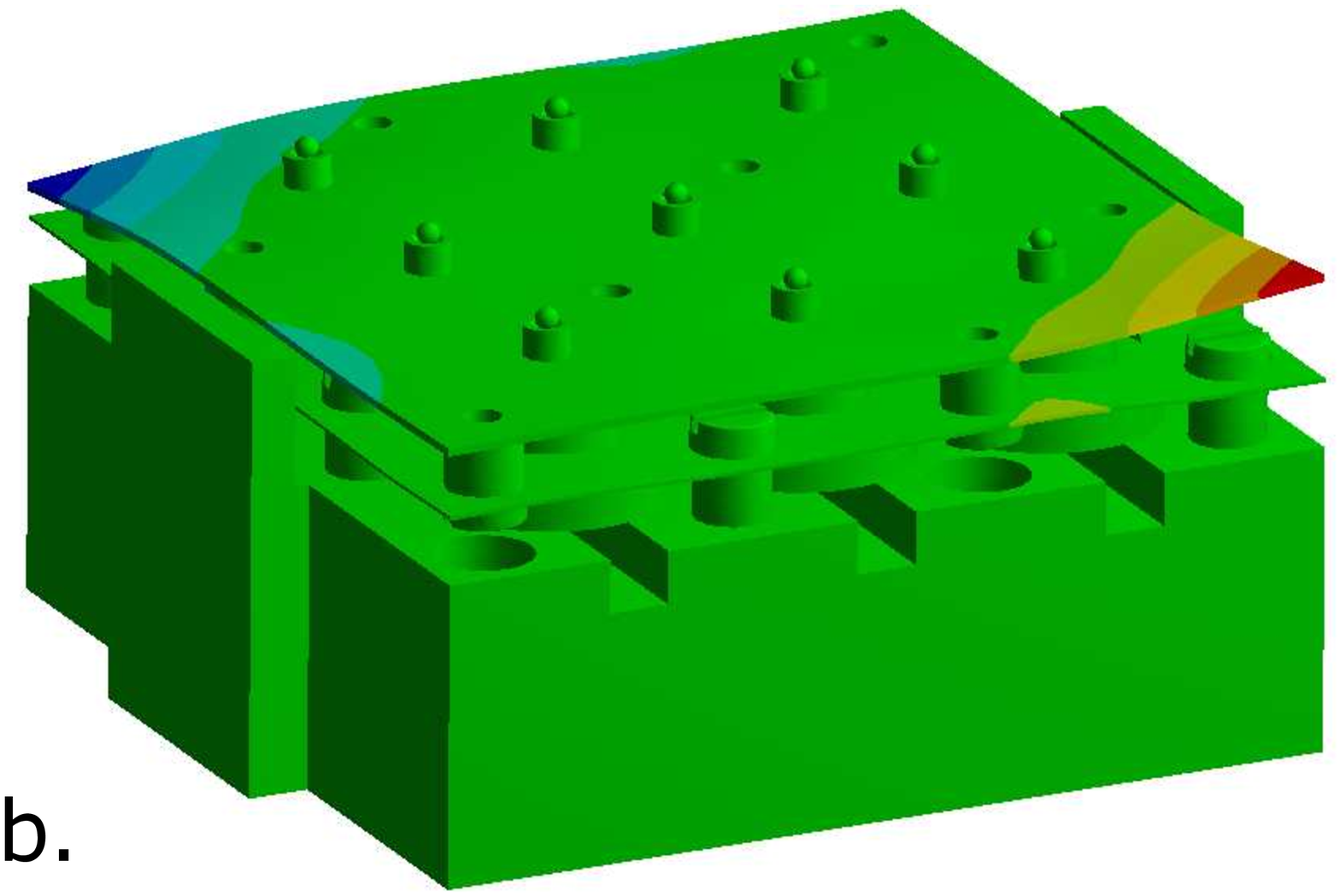}
\caption{Eigenshapes of the first (a, 1556.2\,Hz) and second (b, 1845.6\,Hz) eigenmodes of the slider, as calculated with our FE model. They consist of corner modes in the top and spring plates, i.e. their maximum deflections are observed on two opposite corners of the plates with antisymmetric displacements. The differences between first and second modes are due to the presence or not of screws at the most deformed corners.}
\label{Mode1}
\end{figure}

\begin{figure}[h!]
\centering
\includegraphics[width=0.7\textwidth]{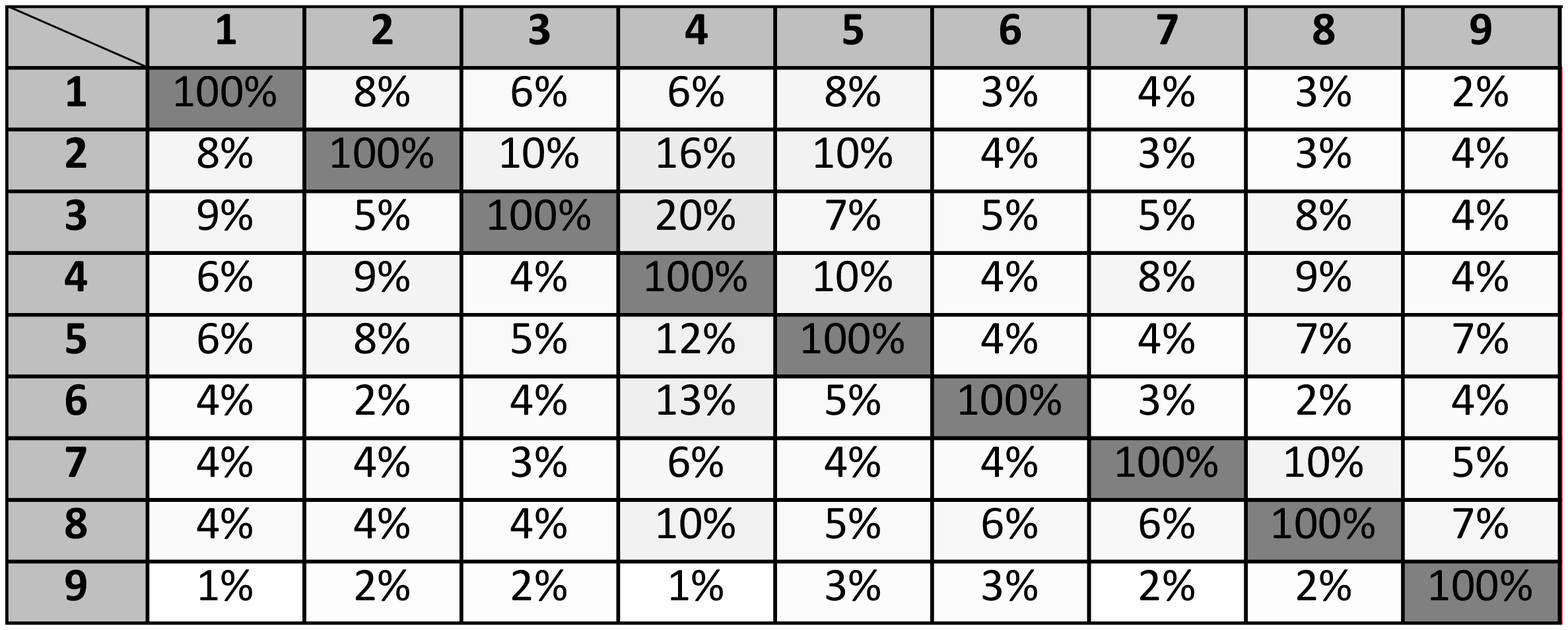}
\caption{Matrix of the ratios $\left|p_{ji}\right|=\left|\frac{a_{ji}}{a_{jj}}\right|$. All values are in \%. The grey level of a box is proportional to its value. Row: label of the impacted asperity. Column: label of the sensor.}
\label{matMax}
\end{figure}

\newpage

\section{Multi-asperity contact model}\label{sec:model}

\begin{figure}[htb!]
\centering
\includegraphics[width=0.8\columnwidth]{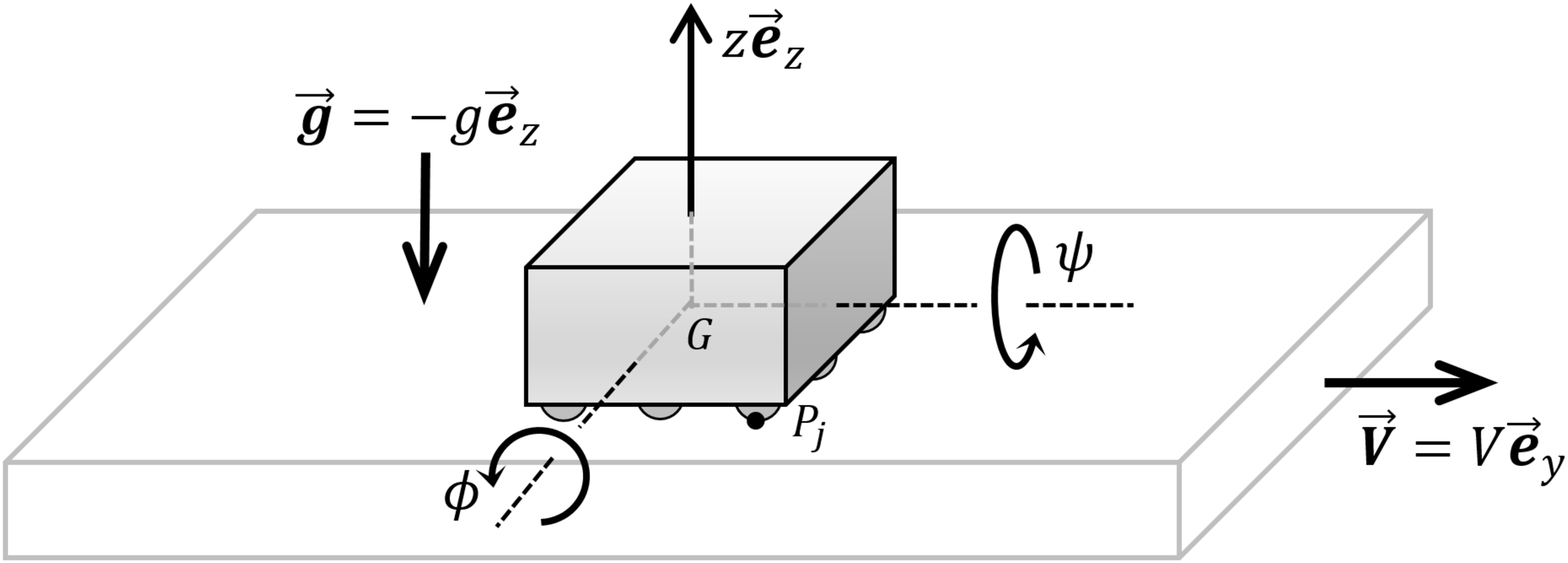}
\caption{Sketch of the multi-asperity contact model.}
\label{Model}
\end{figure}

We model the instrumented slider as a rigid, homogeneous parallelepiped submitted to the gravitational field $\vec{g}=-g\vec{e_z}$ (Fig.~\ref{Model}). We consider only three degrees-of-freedom: the vertical displacement of the centre of mass $G$ of the slider, $z$ along the axis $\vec{e_z}$, and its two rotations around the $G\vec{e_x}$ and $G\vec{e_y}$ axis, respectively $\phi$ and $\psi$. In contrast, the displacement of the centre of mass along the $\vec{e_x}$ and $\vec{e_y}$ axis, as well as the rotation around the $G\vec{e_z}$ axis are constrained. The slider is supported by a track moving along the $\vec{e_y}$ axis at a constant velocity $\vec{V}=V\vec{e_y}$. The track is textured and we assume that the topography is described by the spatial function $h(x,y)$. Interactions between the moving track and slider depend on the contact conditions at the nine hemispheric surface asperities of the slider, the summits of which are located at points $P_j~(j=1,..,9)$.
	
Without loss of generality, the resulting macroscopic force $F_z \vec{e_z}$ and torques $L_x \vec{e_x}$, $L_y \vec{e_y}$ can be obtained from the local normal forces $f_j \vec{e_z}$ (with $f_j\geq0$) exerted by the track on the slider through its surface asperities, and from the distances between the corresponding contact points $P_j$ and the center of mass $G$. Letting

\begin{equation}
	\vec{GP}_j=x_j \vec{e_x}+y_j\vec{e_y}-z_j\vec{e_z}\quad j=1,..,9 \quad \text{and} \quad z_j>0
\end{equation}
leads to the following three equations for the slider's dynamics:
\begin{equation}
\left\{
\begin{aligned}
    m\ddot{z}&=-mg+F_z(f_j)&&=-mg+\sum_{j=1}^9{f_j},\\
    I_x\ddot{\phi}&=L_x(f_j,y_j)&&=+\sum_{j=1}^9{y_j f_j},\\
    I_y\ddot{\psi}&=L_y(f_j,x_j)&&=-\sum_{j=1}^9{x_j f_j},
\end{aligned}
\right.
\label{eq:PFD}
\end{equation}
where $m$, $I_x$ and $I_y$ are, respectively, the mass and moments of inertia around $\vec{e_x}$ and $\vec{e_y}$.

\begin{figure}[htb!]
\centering
\includegraphics[width=0.7\columnwidth]{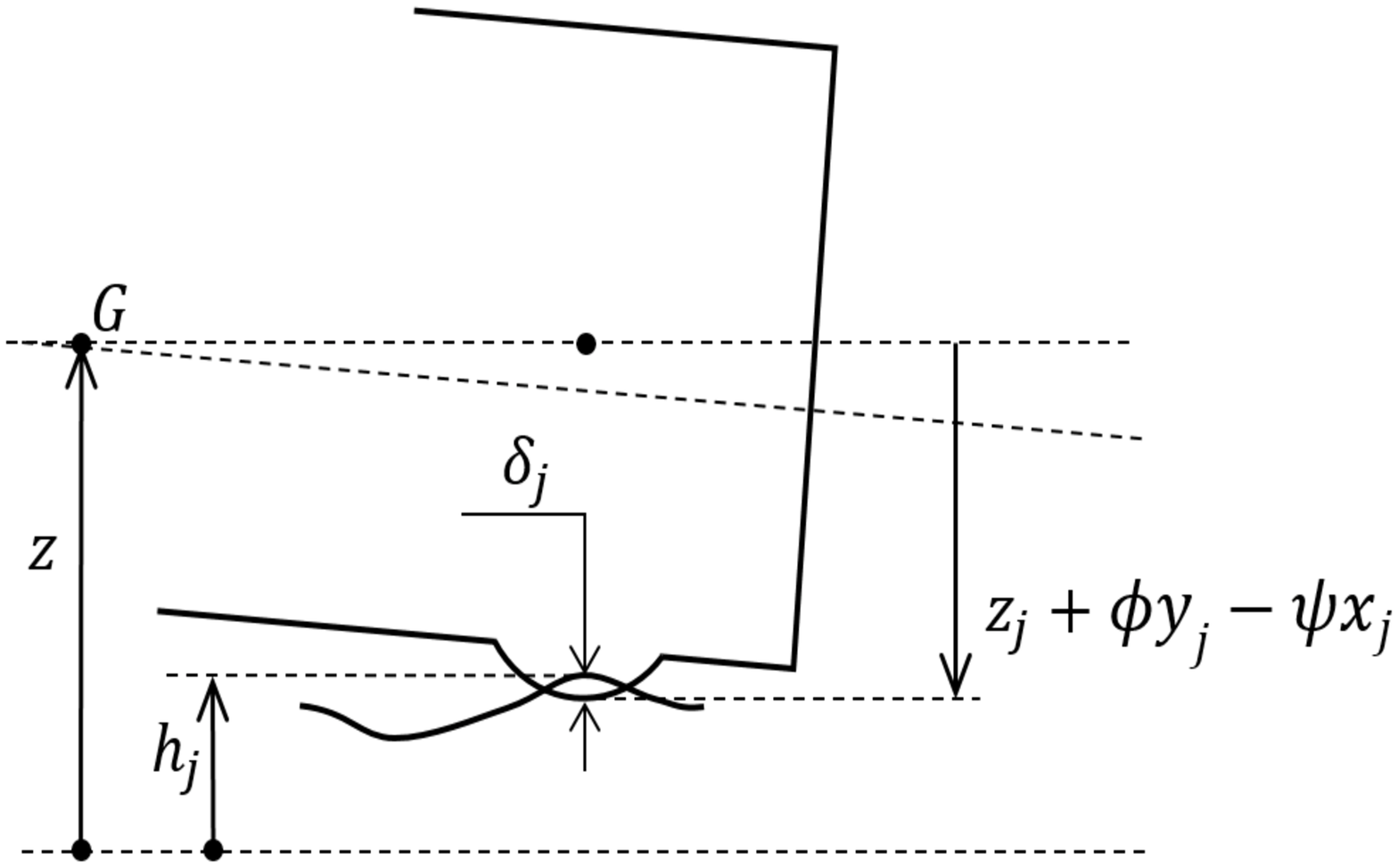}
\caption{Sketch of the normal indentation at contact points. }
\label{Delta}
\end{figure}

To compute the local contact force on asperity $j$, we assume Hertzian behaviour for its elastic part and nonlinear viscous damping for its dissipative part. Thus, we have:
\begin{equation}
    f_j=(1+\alpha_j\dot{\delta}_j)k_j{\delta_j}^{3/2},
    \label{eq:force_contact}
\end{equation}
where $\delta_j$ represents the normal contact indentation (see Fig.~\ref{Delta}), which can be expressed, by choosing convenient origins and under the assumption of small displacement, as follows:
\begin{equation}
    \delta_j=(h_j-z+z_j+\phi y_j-\psi x_j )H(h_j-z+z_j+\phi y_j-\psi x_j),
    \label{eq:delta_full}
\end{equation}
where $h_j=h(x_j,y_j-Vt)$ represents the local altitude of the textured track at the contact point $P_j$, and $H(.)$ is the Heaviside function introduced to account for loss of contact. In practice $k_j$ is obtained from Hertzian theory, so
\begin{equation}
    k_j=k=\frac{4}{3}\sqrt{R} E^*,
\end{equation}
with $R$ the radius of the surface asperity, $E^*$ the composite modulus such that $1/E^* =(1-\nu_1^2)/E_1 +(1-\nu_2^2)/E_2$, with $E_{1,2}$ and $\nu_{1,2}$ the Young moduli and Poisson coefficients of the antagonist solids.

Note that in Eq.~(\ref{eq:force_contact}), the dissipative force is assumed proportional to the elastic force, which we have obtained from  an equivalent viscous damping of the linearised dynamic response around the static deformation. To this end, we assume that (i) the slider's motion is only governed by the vertical displacement $z(\phi=\psi=0)$, (ii) the number of contacts $n$ respects isostatic equilibrium under light load, so that $n$=3, and (iii) the initial distances at the location of the 3 asperities ($h_j+z_j$) are all equal to $D$. Consequently, normal approaches are identical, leading to:
\begin{align}
    \delta_j&=(D-z)H(D-z),\\
    \dot{\delta}_j&=-\dot{z}H(D-z).
\end{align}
By introducing the variable $u=D-z$,$\dot{u}=-\dot{z}$, $\ddot{u}=-\ddot{z}$ and assuming permanent contact $H(u)=1$, we obtain for Eqs.~(\ref{eq:PFD}),~(\ref{eq:force_contact}) and~(\ref{eq:delta_full}):

\begin{align}
    &-m\ddot{u}=-mg+\sum_{i=1}^3{(1+\alpha\dot{u})ku^{3/2}},\\
    &m\ddot{u}+3k(1+\alpha\dot{u})u^{3/2}=mg,\\
    &m\ddot{u}+K(1+\alpha\dot{u})u^{3/2}=N,
\end{align}
with $N=mg$ and $K=3k$. The static equilibrium is easily obtained as $u_s=(N/K)^{2/3}$. Introducing the new variable $q$ defined by $u=u_s(1+\frac{2}{3}q)$, $\dot{u}=\frac{2u_s}{3}\dot{q}$ and $\ddot{u}=\frac{2u_s}{3}\ddot{q}$ leads to:
\begin{equation}
    \frac{2mu_s}{3}\ddot{q}+K{u_s}^{3/2}(1+\frac{2\alpha u_s}{3}\dot{q})(1+\frac{2}{3}q)^{3/2}=N.
\end{equation}
Now we can linearise the motion equation around the static equilibrium $q=0$, that is:
\begin{align}
    (1+\frac{2}{3}\alpha u_s\dot{q})(1+\frac{2}{3}q)^{3/2}&\approx(1+\frac{2\alpha u_s}{3}\dot{q})(1+\frac{2}{3}q)\\
    &\approx1+q+\frac{2\alpha u_s}{3}\dot{q},
\end{align}
and
\begin{equation}
    \frac{2mu_s}{3}\ddot{q}+K{u_s}^{3/2}(1+\frac{2\alpha u_s}{3}\dot{q}+q)=N.
\end{equation}

With $\omega^2=\frac{3K{u_s}^{1/2}}{2m}$,$\tau=\omega t$ and $()'=\frac{d}{d\tau}()$, we get:
\begin{align}
    \frac{d^2q}{\omega^2dt^2}+\frac{2\alpha u_s\omega}{3}\frac{dq}{\omega dt}+q=0,\\
    q''+2\zeta q'+q=0.
\end{align}
So, the coefficient $\alpha$ can be related to an equivalent modal viscous damping $\zeta=\alpha u_s \omega/3$, which gives
\begin{align}
   &\alpha=\frac{3\zeta}{u_s\omega}=\sqrt{6}\zeta K^{-1/2}m^{1/2}u_s^{-5/4}\\
   &\alpha=\sqrt{6}\zeta K^{1/3}m^{-1/3}g^{-5/6}=3^{5/6}\sqrt{2}k^{1/3}m^{-1/3}g^{-5/6}\zeta\\
   &\alpha=(1944k^{2}m^{-2}g^{-5})^{1/6}\zeta\\
\end{align}

Note that the dimension of $\alpha$ has been verified to be $\sim TL^{-1}$ and unit $s.m^{-1}$.

In order to compute the dynamic response of the slider under the excitation of the moving track, a classical numerical time integration method is used, based on the explicit velocity-Verlet scheme. In order to ensure the stability of this time-integration scheme, we must choose a time step $\Delta t$ such as $\Omega\Delta t<2$, where $\Omega$ is the highest natural frequency of the studied dynamic system. In our case, one can estimate this frequency by considering the linearised system around its static equilibrium over $n$ asperities. A typical linearised stiffness $\tilde{k}$ per asperity is given by : 
\begin{equation}
    \tilde{k}=\frac{3}{2}k\delta_s^{1/2}.\label{eq:ktilde}
\end{equation}

With $mg/n=k\delta_s^{3/2}$, Eq.~(\ref{eq:ktilde}) becomes 
\begin{equation}
    \tilde{k}=\frac{3}{2}k\left(\frac{mg}{nk}\right)^{1/3}.
\end{equation}
With $n$ simultaneous asperity contacts, the total linearised stiffness $\tilde{K}$ is
\begin{equation}
    \tilde{K}=\frac{3}{2}(nk)^{2/3}(mg)^{1/3},
\end{equation}
and an estimate of the highest natural frequency $\Omega=\sqrt{\tilde{K}/m}$ is obtained as follows:
\begin{equation}
    \Omega=\frac{3}{2}\left(\frac{nk}{m}\right)^{2/3}g^{1/3},
\end{equation}
leading to $\Delta t<\sqrt{\frac{4m}{\tilde{K}}}=\frac{2\sqrt{2}}{\sqrt{3}}\left(\frac{m}{nk}\right)^{1/3}g^{-1/6}$ which is minimal for $ n=9$. Finally we get a conservative criterion for $\Delta t$ as:
\begin{equation}
    \Delta t<\frac{2\sqrt{2}}{\sqrt{3}}\left(\frac{m}{9k}\right)^{1/3}g^{-1/6}\approx \frac{1}{2}\left(\frac{m}{k}\right)^{1/3}.
\end{equation}

%
\section*{Acknowledgements}
We acknowledge support from Labex CeLyA of Université de Lyon, operated by the French National Research Agency (Grants No. ANR-10-LABX-0060 and No. ANR-11-IDEX-0007). We thank the following colleagues from LTDS for their help: Stéphane Lemahieu (welding and gluing of the piezoelectric sensors), Didier Roux (manufacturing), Matthieu Guibert (design and instrumentation), Youness Benaicha and Karl Landet (finite elements), Davy Dalmas (critical reading of the manuscript).

\section*{CRediT authorship contribution statement}
\JS{\textbf{C. Grégoire:} Investigation, Visualization, Writing - Review \& Editing \textbf{B. Laulagnet:} Conceptualization, Supervision, Writing - Review \& Editing \textbf{J. Perret-Liaudet:} Conceptualization, Supervision, Writing - Review \& Editing \textbf{T. Durand:} Methodology \textbf{M. Collet:} Writing - Review \& Editing \textbf{J. Scheibert:} Conceptualization, Supervision, Writing - Original draft}

\section*{Conflict of interest}
 The authors declare that they have no conflict of interest.


\bibliographystyle{elsarticle-harv}

\end{document}